\documentclass[fleqn,usenatbib]{mnras}

\usepackage{newtxtext,newtxmath}

\usepackage[T1]{fontenc}

\DeclareRobustCommand{\VAN}[3]{#2}
\let\VANthebibliography\thebibliography
\def\thebibliography{\DeclareRobustCommand{\VAN}[3]{##3}\VANthebibliography}

\usepackage{graphicx}	
\usepackage{amsmath}	

\usepackage{CJKutf8}    
\usepackage{multirow}   
\newcommand{\chinese}[1]{\begin{CJK}{UTF8}{bsmi}#1\end{CJK}}

\newcommand{\teff}{\ensuremath{T_{\mathrm{eff}}}}
\newcommand{\feh}{\ensuremath{[\mathrm{Fe/H}]}}
\newcommand{\xfe}{\ensuremath{[\mathrm{X/Fe}]}}
\newcommand{\mgfe}{\ensuremath{[\mathrm{Mg/Fe}]}}
\newcommand{\sife}{\ensuremath{[\mathrm{Si/Fe}]}}
\newcommand{\cafe}{\ensuremath{[\mathrm{Ca/Fe}]}}
\newcommand{\tife}{\ensuremath{[\mathrm{Ti/Fe}]}}
\newcommand{\cfe}{\ensuremath{[\mathrm{C/Fe}]}}
\newcommand{\nife}{\ensuremath{[\mathrm{Ni/Fe}]}}
\newcommand{\crfe}{\ensuremath{[\mathrm{Cr/Fe}]}}
\newcommand{\alphafe}{\ensuremath{[\mathrm{\alpha/Fe}]}}
\newcommand{\dex}{\ensuremath{\,\mathrm{dex}}}
\newcommand{\logg}{\mbox{$\log g$}}

\newcommand{\snrlm}{$\ensuremath{\mathrm{SNR_{\mathrm{LAMOST}}}}$}
\newcommand{\snrmuse}{$\ensuremath{\mathrm{SNR_{\mathrm{MUSE}}}}$}

\newcommand{\ddpayne}{\emph{DD-Payne}}
\newcommand{\ddpayneg}{\emph{DD-Payne-G}}
\newcommand{\ddpaynea}{\emph{DD-Payne-A}}
\newcommand{\ddpayneamp}{\emph{DD-Payne-A-mp}}
\newcommand{\cannon}{\emph{The Cannon}}
\newcommand{\thepayne}{\emph{The Payne}}
\newcommand{\astronn}{\emph{Astro-NN}}
\newcommand{\pampelmuse}{PampelMUSE}
\newcommand{\crlb}{\emph{Chem-I-Calc}}
\newcommand{\abinitio}{\emph{ab initio}}

\newcommand{\Rsun}{$R_{\odot}$}
\newcommand{\kms}{{\rm km s}$^{-1}$}
\newcommand{\rgc}{$R_{\mathrm{gc}}$}

\newcommand{\vmic}{$V_\mathrm{mic}$}
\newcommand{\perpixel}{$\text{pix}^{-1}$}

\title[MUSE Stellar labels from Data-Driven Payne]{Reliable stellar abundances of individual stars with the MUSE integral-field spectrograph}

\author[Zixian Wang et al.]{
Zixian Wang(\chinese{王梓先}),$^{1,2}$\thanks{E-mail: zwan0382@uni.sydney.edu.au/wang.zixian.astro@gmail.com}
Michael R. Hayden,$^{1,2}$
Sanjib Sharma,$^{1,2}$
Maosheng Xiang(\chinese{向茂盛}),$^{4}$
\newauthor
Yuan-Sen Ting(\chinese{丁源森}),$^{5,6,7,8,9}$
Joss Bland-Hawthorn,$^{1,2,3}$
Boquan Chen$^{1,2}$
\\
$^{1}$Sydney Institute for Astronomy, School of Physics, A28, The University of Sydney, NSW, 2006, Australia\\
$^{2}$ARC Centre of Excellence for All Sky Astrophysics in Three Dimensions (ASTRO-3D)\\
$^{3}$Miller Professor, Miller Institute, UC Berkeley, Berkeley CA 94720\\
$^{4}$Max-Planck Institute for Astronomy, Königstuhl 17, D-69117 Heidelberg, Germany; mxiang@mpia.de \\
$^{5}$Research School of Computer Science, Australian National University, Acton ACT 2601, Australia \\
$^{6}$Research School of Astronomy and Astrophysics, Australian National University, ACT 2611, Australia \\
$^{7}$Institute for Advanced Study, Princeton, NJ 08540, USA \\
$^{8}$Department of Astrophysical Sciences, Princeton University, Princeton, NJ 08544, USA \\
$^{9}$Observatories of the Carnegie Institution of Washington, 813 Santa Barbara Street, Pasadena, CA 91101, USA \\
}

\date{Accepted XXX. Received YYY; in original form ZZZ}

\pubyear{2021}

\DeclareUnicodeCharacter{2212}{-}

\begin{document}
\label{firstpage}
\pagerange{\pageref{firstpage}--\pageref{lastpage}}
\maketitle

\begin{abstract}
We present a novel approach to deriving stellar labels for stars observed in MUSE fields making use of data-driven machine learning methods.
Taking advantage of the comparable spectral properties (resolution, wavelength coverage) of the LAMOST and MUSE instruments, we adopt the Data-Driven Payne (\ddpayne{}) model used on LAMOST observations and apply it to stars observed in MUSE fields.
Remarkably, in spite of instrumental differences, according to the cross-validation of 27 LAMOST-MUSE common stars, we are able to determine stellar labels with precision better than 75K in \teff{}, 0.15 dex in \logg{}, and 0.1 dex in abundances of \feh{}, \mgfe{}, \sife{}, \tife{}, \cfe{}, \nife{} and \crfe{} for current MUSE observations over a parameter range of $3800<\teff{}<7000$ K, $-1.5<\feh{}<0.5$ dex. 
To date, MUSE has been used to target 13,000 fields across the southern sky since it was first commissioned six years ago and it is unique in its ability to study dense star fields such as globular clusters or the Milky Way bulge.
Our method will enable the automated determination of stellar parameters for all stars in these fields. Additionally, it opens the door for applications to data collected by other spectrographs having resolution similar to LAMOST. With the upcoming BlueMUSE and MAVIS, we will gain access to a whole new range of chemical abundances with higher precision, especially critical s-process elements such as [Y/Fe] and [Ba/Fe] that provide key age diagnostics for stellar targets. 
\end{abstract}




\section{Introduction}
\label{s:intro}

The formation and evolution of the Milky Way is one of the outstanding questions facing astrophysics today. 
The study of stars and stellar populations is a pillar of Galactic Archaeology \citep{2002ARA&A..40..487F}, as they contain the chemical imprint of the gas from which they formed and serve as birth tags.
This allows stars to be used as a fossil record to unravel the history of the Milky Way.
During the past decade, traditional fiber-fed spectrographs employed by most large-scale spectroscopic surveys like RAVE \citep{2006AJ....132.1645S}, LAMOST \citep{2012RAA....12..735D, 2012RAA....12..723Z}, Gaia-ESO \citep{2012Msngr.147...25G}, GALAH \citep{2015MNRAS.449.2604D}, and APOGEE \citep{2017AJ....154...94M}
combined with astrometric and photometric information from Gaia \citep{2018A&A...616A...1G, 2021A&A...649A...1G} provided detailed chemodynamics for millions of stars from the solar neighborhood to even  several kilo-parsecs away \citep{2016ARA&A..54..529B}. 

The large volume of spectroscopic observations is posing great challenges for data analysis, particularly in deriving stellar labels (atmospheric parameters and chemical abundances) precisely and efficiently from spectra. 
In the past few years, with the use of radiation transfer codes (e.g. SME \citealt{1996A&AS..118..595V}, Turbospectrum \citealt{1998A&A...330.1109A, 2012ascl.soft05004P}, MOOG \citealt{2012ascl.soft02009S}), many model-driven (e.g. \thepayne{} \citealt{2019ApJ...879...69T} and \emph{StarNet} \citealt{2018MNRAS.475.2978F, 2020MNRAS.498.3817B}) and data-driven methods (e.g. \cannon{} \citealt{2015ApJ...808...16N} and \astronn{} \citealt{2019MNRAS.483.3255L}) and models in between (e.g. \ddpayne{} \citealt{2019ApJS..245...34X} and \emph{Cycle-StarNet} \citealt{2021ApJ...906..130O}) have been developed for this purpose.
One benefit of the data-driven approach is it can generate a model from a training set sample, rather than traditional physics-based approaches using theoretical stellar atmospheres; this 
is much easier to do and in general leads to more precise stellar parameters as it can potentially make use of the full wavelength region of the spectra. 

However, one limitation of the data-driven methods is that some chemical abundances could be determined by astrophysical correlations in the training set, instead of physically motivated measurements from spectral features. 
This has a non-negligible impact on the precision of some stellar labels, especially those derived from low-resolution spectra, where many of the absorption features are blended.

To overcome this issue, \citealt{2017ApJ...849L...9T, 2018ApJ...860..159T} used the methods outlined in \thepayne{} \citep{2019ApJ...879...69T} to build a neural network to determine stellar labels. They imported a new term in the loss function which takes into account the similarity of gradient spectra from the data-driven model and Kurucz model atmospheres \citep{1970SAOSR.309.....K, 1993KurCD..18.....K, 2005MSAIS...8...14K}, to regularize the training process and force the model to measure labels from real absorption features. They verified that more than 10 elements can be extracted even from the low-resolution spectra as in LAMOST.
Based on this approach \citep{2017ApJ...849L...9T}, \citealt{2019ApJS..245...34X} further developed the Data-Driven Payne (hereafter \ddpayne{}) which used input training labels from the high-resolution spectroscopic surveys APOGEE \citep{2019ApJ...879...69T} and GALAH \citep{2018MNRAS.478.4513B} to determine stellar labels for LAMOST DR5 spectra. 
These authors found that they could precisely estimate close to 20 stellar labels for more than 6 million spectra. These studies highlight  
that if one has a reliable training set then one can obtain precise elemental abundances even from low-resolution spectra.

The methods outlined above demonstrate how one can measure stellar labels for millions of spectra using machine learning approaches. However, all of the surveys described previously are conducted using fiber-fed spectrographs, which take spectra by plugging a series of optical fibers onto a plate based on the positions of stars on the sky.
Fibers are typically set to have a minimum separation (e.g. $30^{\prime \prime}$ for GALAH; $56-71.5^{\prime \prime}$ for APOGEE; $268.2^{\prime \prime}$ for LAMOST) during each observation. And each fiber has a typical diameter of $2\sim3^{\prime \prime}$. 
This means that these instruments can not observe dense fields such as cores of globular clusters and the very dense regions of Galactic bulge, where the separation of stars is less than $1^{\prime \prime}$.
However, stars in these fields also play an essential role in the Galactic evolution, it is vital to obtain more complete spectroscopic observations in these regions to unravel the history of the Milky Way.

The Multi-Unit Spectroscopic Explorer (MUSE) \citep{2010SPIE.7735E..08B} on the Very Large Telescope is an integral field spectrograph, which is regularly used to observe external galaxies and can be an ideal solution to deal with high stellar density fields. 
With the advanced development of image slicer and spectrometer, MUSE can perform both imaging and spectroscopy simultaneously. 
The benefit is that each MUSE spectrum is obtained for a spatial pixel ($0.2^{\prime \prime} \times 0.2^{\prime \prime}$) on the sky rather than the area of the fiber, so it is not restricted by the fiber size or collision effects.
The data format of MUSE is a 3-D data-cube, with 2D in spatial on the sky and 1D in wavelength. Combining with the PSF-fitting spectral extraction software \pampelmuse{} \citep{2013A&A...549A..71K}, MUSE becomes the best instrument to obtain spectra in fields with high stellar density.

There have been several studies on stars in globular clusters using spectra from MUSE observations \citep{2016A&A...588A.148H, 2016A&A...588A.149K}. 
Near the core of these clusters, more than 6000 spectra (e.g., 47 Tuc at a distance of 4 kpc analyzed by \citealt{2016A&A...588A.149K}) can be extracted on average per MUSE cube (in wide-field mode with a field of view of $1^{\prime} \times 1^{\prime}$). Currently, most studies focused exclusively on kinematics (e.g. \citealt{2016A&A...588A.149K, 2018MNRAS.473.5591K, 2018A&A...616A..83V}). 
Some studies explored specific spectral region or metallicity (e.g. \citealt{2019A&A...631A.118G, 2020A&A...635A.114H}).
However, these spectra contain a wealth of chemical information that has not yet been fully explored. 
This is because currently there is no robust and easy-to-use method to automatically derive chemical abundances for stars in these fields. 

The success of data-driven models in 
deriving about $20$ abundances from LAMOST spectra highlights the potential of other 
low-resolution surveys, for example MUSE,  
having wavelength coverage and resolution 
similar to LAMOST to also derive abundances.
Inspired by this, in this paper, we develop an approach based on the \ddpayne{} \citep{2017ApJ...849L...9T, 2019ApJS..245...34X}, to measure stellar labels from MUSE observations.
Using this approach we derive robust stellar labels, including effective temperature (\teff{}), surface gravity (\logg{}), metallicity \feh{} and abundances of four $\alpha$ elements, \mgfe{}, \sife{}, \cafe{} and \tife{}.

In Section~\ref{s:data}, we introduce the data-sets used in our analysis and the spectral extraction method used by us. In Section~\ref{s:meth}, we outline the procedures used to measure stellar labels for MUSE spectra. 
We do two validation tests in Section~\ref{s:resu}. First, we verify the bias and dispersion of our approach by applying it to common stars between LAMOST and MUSE. Second, for dense fields, we explore the \feh{}-\mgfe{} distribution of stars towards the bulge region using MUSE observations and compare it with that obtained from the high-resolution APOGEE survey. In Section~\ref{s:predict}, we explore and discuss the dependence of label precision on magnitude and exposure time. In Section~\ref{s:discuss}, we discuss the exciting prospects of using MUSE to observe dense stellar fields in the Milky Way and other galaxies. Finally, a summary of our results and conclusions is presented in Section~\ref{s:summary}.


\section{Data}
\label{s:data}

\subsection{MUSE}
\label{s:data-muse}

\subsubsection{Overview}
\label{ss:data-muse-ow}

MUSE (the Multi-Unit Spectroscopic Explorer) \citep{2010SPIE.7735E..08B, 2014Msngr.157...13B} is an optical wide-field integral field spectrograph using the image slicing technique mounted on the UT4 of the Very Large Telescope at the Paranal Observatory in Chile. 
MUSE operates in two spatial modes, the wide-field mode (WFM) and the narrow-field mode (NFM).
The wide-field mode covers a FoV of $1^{\prime} \times 1^{\prime}$ with  $0.2^{\prime \prime} \times 0.2^{\prime \prime}$ spatial bins. 
The narrow-field mode covers a FoV of $7.5^{\prime \prime} \times 7.5^{\prime \prime}$ with  $0.025^{\prime \prime} \times 0.025^{\prime \prime}$ spatial bins. 
It operates in two wavelength modes, the nominal mode covering 4800-9300~$\Angstrom$ and the extended mode covering 4650-9300~$\Angstrom$. It samples the wavelength in 1.25$\Angstrom$ bins at a spectral resolution of R$\sim$3000.
To date, more than 13000 data-cubes have been observed by MUSE and they are publicly available via the ESO Science Portal Website\footnote{\url{http://archive.eso.org/scienceportal/home}}. MUSE is typically used to conduct studies on external galaxies but it has also been used to study a wide variety of other things, e.g., exoplanets \citep{2018Icar..302..426I, 2019NatAs...3..749H}, Galactic regions \citep{2015A&A...582A.114W, 2015MNRAS.450.1057M} and resolved stellar populations of Globular clusters \citep{2016A&A...588A.148H, 2018MNRAS.473.5591K}.

\subsubsection{Reductions}
\label{ss:data-muse-rd}
The data format of MUSE observations follows the ESO science data product standard for data-cubes. Each data-cube has three dimensions, two in spatial and one in spectral. 
The data processing pipeline transforms the raw CCD-based data into fully calibrated and corrected data-cubes, which can be directly used for studies. The transformation processes include bias, dark and flat-field reduction, wavelength and flux calibration, line spread function and illumination correction, sky creation, and astrometrically correlation and multiple exposures combination, which are discussed in detail in \citealt{2012SPIE.8451E..0BW, 2014ASPC..485..451W, 2020A&A...641A..28W}. 
For this study, all the data-cubes we use are published online and have been processed and calibrated using the MUSE Data Processing Pipeline\footnote{\url{https://www.eso.org/sci/software/pipelines/muse/}}.

\subsubsection{Stellar Spectral Extraction}
\label{ss:data-muse-se}

To extract stellar spectra, we use the \pampelmuse{} package \citep{2013A&A...549A..71K} which was specifically developed for 3D data-cubes in dense fields. It assumes the data-cube to be a sum of many overlapping PSFs and sky background. 
First, a single object is represented by a PSF function such as Moffat and Gaussian, and assuming parameters in the PSF function change smoothly with wavelengths. 
Next, the PSF photometry is employed to fit each star in each wavelength layer and then all the layers are combined to generate a single stellar spectrum. 
One great advantage of this method is that it can de-blend the spectra of different stars. 
Therefore, MUSE observations combined with \pampelmuse{} become an ideal way to study dense fields.

To get spectra in the field using \pampelmuse{}, a reference catalog with stellar positions and magnitude is needed as input. 
Various reference catalogs can be used, such as all-sky surveys (e.g. Gaia \citealt{2018A&A...616A...1G, 2021A&A...649A...1G}, Sky-Mapper \citealt{2007PASA...24....1K} or HST) or smaller surveys for specific purposes. In this study, we mainly use LAMOST DR5 and VVV DR2 \citep{2017yCat.2348....0M} as reference catalogs for the two validation test data, respectively.
Moreover, a transformation function of the filter band is needed when aligning the MUSE data-cube slice with reference catalogs. 
Here we refer to the transformation function from SVO Filter Profile Service\footnote{\url{https://svo.cab.inta-csic.es/main/index.php}} and use the Moffat function as the PSF profile, which has better performance than the Gaussian profile when fitting flux distribution of stars (see Figure 3 in \citealt{2013A&A...549A..71K}). 

After going through all processes in \pampelmuse{}, including source position transformation, PSF fitting, sky background estimation and subtraction, and unresolved star reduction, the output is a list of stellar spectra which are identified as resolvable by the algorithm. 
A few additional cleaning steps (removal of emission lines and non-stellar features) are performed on the spectra and will be discussed in Section~\ref{sss:resu-bulge-data}.

\subsection{LAMOST}
\label{s:data-lamost}

\subsubsection{Overview}
\label{ss:data-lamost-ow}

The LAMOST Galactic survey \citep{2012RAA....12..735D, 2012RAA....12..723Z, Liu_2013, 2015RAA....15.1089L} is the first dedicated spectroscopic survey to obtain spectra of millions of stars covering most of the sky. 
LAMOST DR5 has released more than 8 million stellar spectra covering the optical wavelength regime 3700-9000 $\Angstrom$ with low-resolution R$\sim$1800. 
For about 5 million spectra, LAMOST DR5 also provides the basic stellar labels such as \teff{}, \logg{} and \feh{} which were derived with the LAMOST stellar parameter pipeline (LASP; \citealt{2011RAA....11..924W, 2015RAA....15.1095L}). There are also several value-added catalogs providing stellar labels with some chemical abundances and extinction via different methods (e.g., \citealt{2015MNRAS.448..822X, 2016RAA....16..110L, 2017MNRAS.464.3657X}). Despite LAMOST's low spectral resolution, \citealt{2017ApJ...849L...9T, 2018ApJ...860..159T} has verified that $\geq$10 chemical abundances should be derivable with precision on the level of $0.1\sim0.2$ dex or better.

\subsubsection{Data-Driven Payne}
\label{ss:data-lamost-ddp}

Given the success of machine learning and data-driven models in stellar parameter determination, an effective way to estimate stellar labels from MUSE spectra would be to develop a data-driven model for the MUSE spectra based on stellar labels from other high-resolution spectroscopic surveys. 
Presently, we do not have enough common stars (ideally $\sim10^4$) covering a wide range in the parameter space between MUSE and other surveys for a robust training set. 
However, with LAMOST spectra having similar resolution and wavelength range as MUSE (see Table~\ref{tab:tab1}), we can apply a model developed for LAMOST to MUSE spectra. 
\citealt{2017ApJ...849L...9T, 2019ApJS..245...34X} (hereafter T17 \& X19) had developed a \ddpayne{} model for LAMOST, which we discuss below. 
It was shown to have good performance and hence we adopted it for our analysis.

\ddpayne{} utilizes the neural network interpolator and the fitting technique from \thepayne{} \citep{2019ApJ...879...69T} combined with physical gradient spectra from Kurucz spectral model \citep{1970SAOSR.309.....K, 1993KurCD..18.....K, 2005MSAIS...8...14K} to regularize the training process.
This model can derive stellar labels of \teff{}, \logg{}, micro-turbulence velocity ($V_{\rm mic}$) and 16 reliable chemical abundances.
Compared with other similar models (\citealt{2017ApJ...836....5H, 2020ApJ...898...58W}), the biggest advantage of \ddpayne{} is it can enforce these stellar labels to be measured physically from the spectral features.
For $\rm{SNR}_{\rm{LAMOST}}^{\rm{g-band}}>50$ \perpixel{}, the typical theoretical precision of the \ddpayne{} abundances is $\sim$0.05 dex for Fe, Mg, Ca, Ti, Cr and Ni, $\sim$0.1 dex for C, N, O, Na, Al, Si, Mn, and Co. This study demonstrates that while obtaining reliable elemental abundances remains challenging for low-resolution spectra, precise abundances are still derivable for spectra with SNR$\sim$50 \perpixel{}. 

\begin{table}
    \caption{Comparison of MUSE and LAMOST survey.}
    \label{tab:tab1}
        \begin{tabular}{lcc}
            \hline
            Survey & Wavelength range ($\Angstrom$) & Resolution\\
            \hline
            MUSE & 4750-9300 & R$\sim$3000\\
            LAMOST & 3700-9000 & R$\sim$1800\\
            \hline
        \end{tabular}
\end{table}


\subsection{Validation data and the motivations}
\label{s:data-vbd}

We do two different validation tests.  
Since the model used by us is designed for LAMOST spectra, the question is how precisely does it work when applied to MUSE spectra? Therefore, for the first validation test, we run the model on both the LAMOST spectra and the MUSE spectra of the same  star and compare the results. 
We cross-matched the LAMOST DR5 catalog from  X19\footnote{\url{http://dr5.lamost.org/doc/vac}} with all the MUSE observations so far published on the ESO website, and found 79 stars across 162 MUSE data-cubes.

\begin{figure*}
\includegraphics[width=1.48\columnwidth]{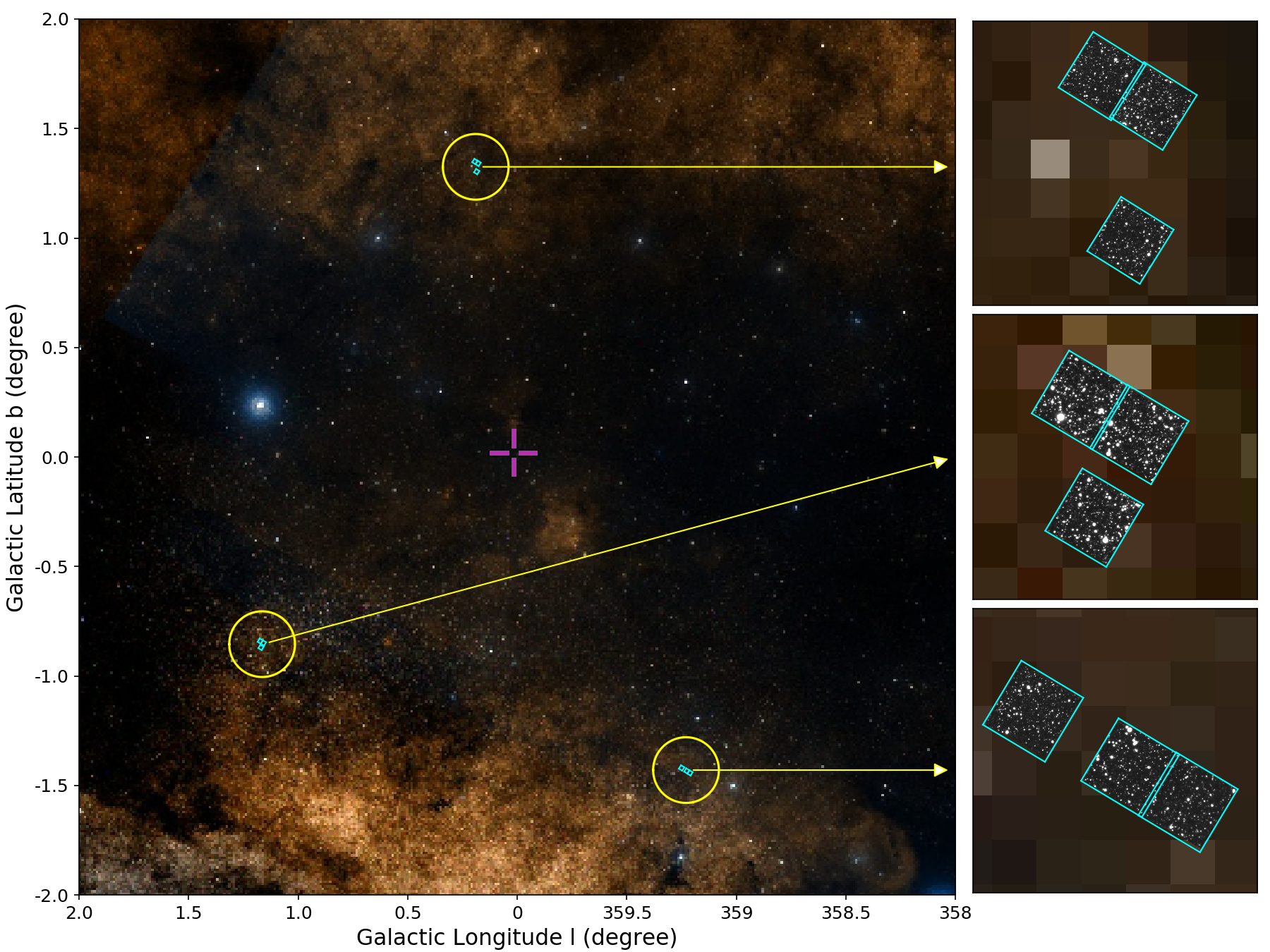}
\caption{
Distribution of 9 fields observed by MUSE (0101.B-0381(A), PI: Zoccali) in the Galactic coordinate with the background taken by DSS2. The nine square fields are marked in cyan and are divided into three areas. The purple cross in the middle represents the Galactic center where $(l, b)=(0, 0)$. On the right-hand side are zoomed-in figures of these three areas guided by yellow arrows. The white light picture of the MUSE data-cube is shown in each square field.
}
\label{f:bulge_cube}
\end{figure*}

The second validation is done by comparing 
the abundance distribution of stars in the inner Galaxy (bulge region) with those obtained from the high-resolution APOGEE survey. 
The Galactic bulge region has a very high stellar density and its stars suffer from strong extinction. 
Hence, validation of bulge stars is important to test the performance of our method in challenging conditions.
Therefore, for the second test, we selected 29 data-cubes (0101.B-0381(A), PI: Zoccali) covering nine fields that are less than 3 degrees to the Galactic center.
They are dense fields and provide a large sample of stars for which to verify our spectroscopic abundances. Figure \ref{f:bulge_cube} shows the distribution of 9 fields in the Galactic coordinate with the background taken by DSS2. The nine square fields are marked in cyan and are divided into three areas. 
The purple cross in the middle represents the Galactic center where $(l,b)=(0,0)$. 
On the right-hand side are zoom-in figures of these three areas guided by yellow arrows. The white light picture of the MUSE data-cube is shown in each square field. 
For each cube field, there are three or four repetitive exposures.
Galactic bulge fields present some unique challenges and need some additional reprocessing, such as low SNR wavelength cut-off, emission line masking, etc, which will be discussed in detail in Section~\ref{sss:resu-bulge-data}.

\section{Methods}
\label{s:meth}

The feasibility of applying the \ddpayne{} on MUSE observations is due to the similar wavelength range and spectral resolution of the MUSE and LAMOST spectrographs.
The processes of deriving stellar labels from MUSE observations include five steps:
\begin{itemize}
 \item Stellar spectra are extracted from MUSE data-cubes by running the automatic spectra extraction algorithm \pampelmuse{} \citep{2013A&A...549A..71K} to obtain spectra at R$\sim$3000. This step was talked about in Section~\ref{ss:data-muse-se}.
 \item The R$\sim$3000 MUSE spectra are degraded by using the line spread function (LSF) of these two instruments to R$\sim$1800.
 \item The spectra are normalized by a pseudo-continuum derived in the same way as in T17 \& X19, to make 
 it compatible with the LAMOST \ddpayne{} model. 
 \item The degraded and normalized spectra are then fitted by three  \ddpayne{} models of X19 to derive stellar labels.
 \item For each label, the cross-correlation of gradient spectra between \ddpayne{} and Kurucz spectral model \citep{1970SAOSR.309.....K, 1993KurCD..18.....K, 2005MSAIS...8...14K} are calculated to check if the label is estimated from real physical features rather than astrophysical correlations.
 \item All the stellar information is generated and combined to create an output catalog.
\end{itemize}
The detailed process of each step and the calculating procedure is described in the following sub-sections.

\subsection{Degrading and normalizing the MUSE spectra}
\label{ss:meth-degrade}

To make \ddpayne{} executable on MUSE spectra, we need to make the MUSE spectra as similar to LAMOST spectra in its wavelength grid and spectral resolution as possible, which leads to the process of degrading and interpolating MUSE spectra.
This is the procedure that determines whether we can derive precise stellar labels from MUSE spectra successfully.

To minimize the difference between LAMOST and MUSE, we employ the line spread function (LSF) of both instruments from X19 and \citealt{2017A&A...608A...1B}. The line spread function is FWHM in each wavelength grid. The process of degrading is performed by Gaussian kernel smoothing over the whole wavelength with $\sigma\mathrm{(\lambda_{i})}$, which is calculated from the equation given by
\begin{equation}
\sigma\mathrm{(\lambda_{i})}=\sqrt{\left(\frac{\mathrm{FWHM_{L}(\lambda_{i})}}{2.355}\right)^{2}-\left(\frac{\mathrm{FWHM_{M}(\lambda_{i})}}{2.355}\right)^{2}},
\label{e:eqn1}
\end{equation}
where $\mathrm{FWHM_{L}(\lambda_{i})}$ and $\mathrm{FWHM_{M}(\lambda_{i})}$ are the LAMOST and MUSE line-spread function, respectively.
After degrading, the spectra will be normalized by a pseudo-continuum which is derived by smoothing the spectra with a Gaussian kernel of 50$\Angstrom$ in width, which is the same method employed by T17 \& X19 for \ddpayne{} training set spectra normalization. This is also an important procedure to ensure the similarity between these two sets of spectra.

Figure \ref{f:lm_muse_comparison} shows the normalized spectra of one single star observed by both LAMOST (in orange) and MUSE (in blue) and their residuals. The MUSE spectra were degraded using the above method. 
The gray areas are the masked areas due to telluric lines and other interference in the LAMOST wavelength grid, which will not be  processed by \ddpayne{}.
This figure demonstrates that for the same star, the difference between the LAMOST and the MUSE spectra is less than 0.03 even in regions with strong absorption lines. This is encouraging and is essential for reliably estimating chemical abundances.
For the LAMOST-MUSE cross-validation data-cubes, the spectral SNR \perpixel{} is in general higher than 200, which is much more than the average SNR of spectra in LAMOST ($\sim$40). After degrading (binning), the SNR is expected to increase even further.

\subsection{Estimating stellar labels and errors}
\label{ss:meth-ddp}

\begin{figure*}
\includegraphics[width=1.86\columnwidth]{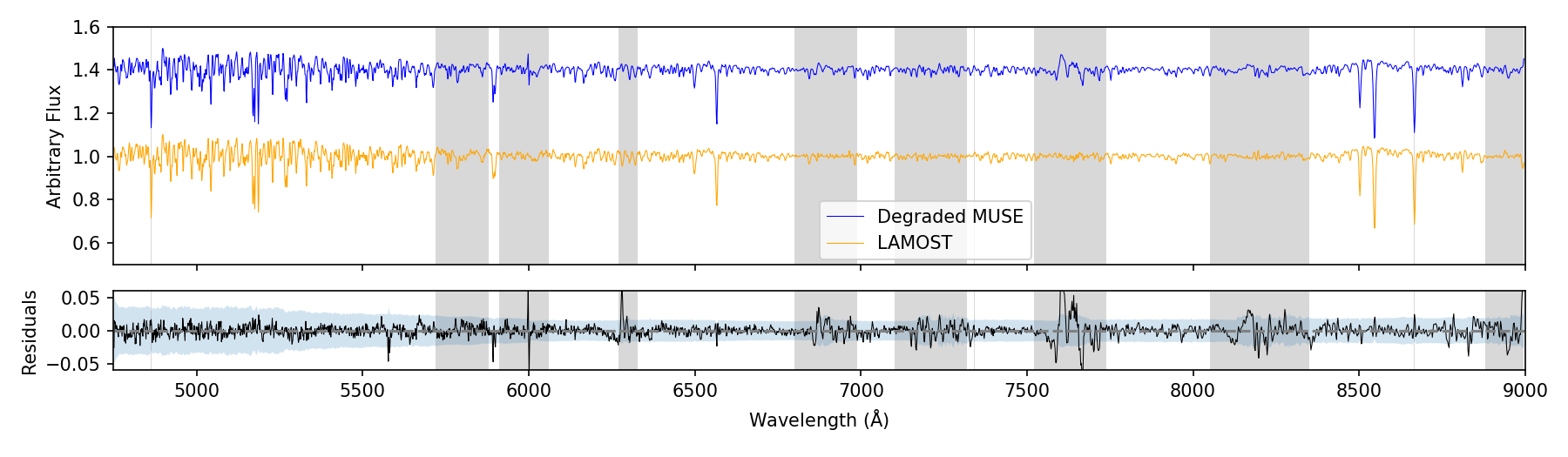}
\caption{
Comparison of LAMOST spectra (in yellow) with MUSE spectra (in blue) of the same star. The MUSE spectra have been degraded to the same resolution. Then both spectra were normalized in the same way as X19 with a pseudo-continuum derived by smoothing the spectra with a Gaussian kernel of 50$\Angstrom$ in width. Gray areas are the masked pixels (due to telluric lines and other sources of interference) which will not be used when fitting with \ddpayne{}. The bottom plot shows the residuals of these two spectra. From the residuals, we can see that LAMOST spectra and MUSE spectra are similar with residuals less than 0.03. The blue shaded area represents the uncertainty of the residual, which is calculated using the flux error of the LAMOST and MUSE spectra. The flux error of MUSE spectra is re-estimated by multiplying $\sigma$ in the distribution of flux residuals of the fitting divided by the flux error, see the right panel of Figure~\ref{f:muse_ddp_comparison}}.
\label{f:lm_muse_comparison}
\end{figure*}

\begin{figure*}
\includegraphics[width=1.86\columnwidth]{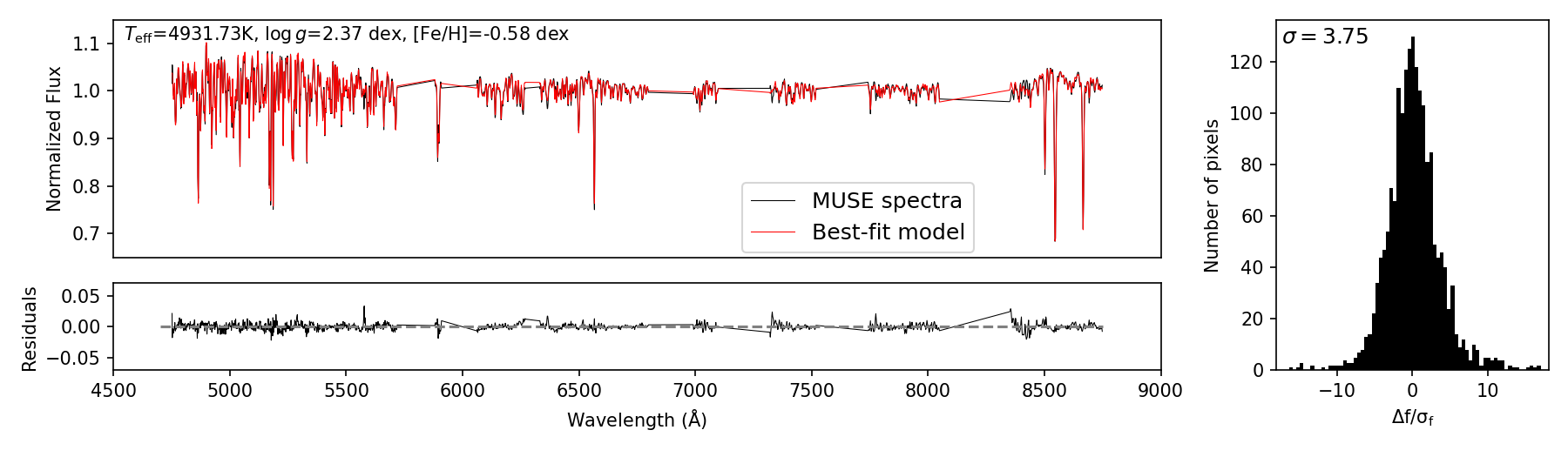}  
\caption{
Comparison of MUSE spectra (in black) with model predicted spectra (in red) from \ddpayneg{} on the same star as Figure \ref{f:lm_muse_comparison}. The left bottom plot shows their residuals. The masked regions in grey in Figure \ref{f:lm_muse_comparison} are not shown in this plot. From the residuals we can see the difference between MUSE spectra and predicted spectra is small, with the residuals less than 0.03 even in absorption regions, demonstrating that \ddpayne{} training models can fit the MUSE spectra very well. The right panel shows the distribution of flux residuals divided by the flux error. A Gaussian fit to the distribution yields a standard deviation of 3.75, which demonstrates the underestimation of flux error of MUSE spectra.
}
\label{f:muse_ddp_comparison}
\end{figure*}

After degrading and normalizing the MUSE spectra to the LAMOST wavelength grid and spectral resolution, the next step is to estimate the stellar labels using the \ddpayne{} models.
There are three models (so-called \ddpayneg{},  \ddpaynea{} and \ddpayneamp{} below) trained by X19 with different training sets.
The \ddpayneg{} model used spectra from LAMOST ($\rm{SNR}_{\rm{LAMOST}}^{\rm{g-band}}>50$ \perpixel{} and $\rm{SNR}_{\rm{LAMOST}}^{\rm{g-band}}>30$ \perpixel{} for stars with $\feh{}<-0.6$) and stellar labels from GALAH DR2 \citep{2018MNRAS.478.4513B} while the \ddpaynea{} model adopted LAMOST spectra ($\rm{SNR}_{\rm{LAMOST}}^{\rm{g-band}}>50$ \perpixel{}) and stellar labels from APOGEE-\emph{Payne} \citep{2019ApJ...879...69T} for their respective cross-matched common samples.
The \ddpayneg{} derives stellar labels for \teff{}, \logg{}, \feh{} and abundance ratios [X/Fe] for 20 elements: Li, C, O, Na, Mg, Al, Si, Ca, Ti, V, Cr, Mn, Co, Ni, Cu, Zn, Y, Ba, and Eu. \ddpaynea{} derive stellar labels of \teff{}, \logg{}, micro-turbulence velocity (\vmic), \feh{} and abundance ratios for 15 elements: C, N, O, Na, Mg, Al, Si, Ca, Ti, Cr, Mn, Co, Ni, Cu, and Ba. 
Since X19 described these models in great detail, we direct the readers to this paper for more information about the selection criteria, validation and comparisons of the performances of different models.
However, we plot the parameter ranges of these training sets in \teff{}, \logg{} and \feh{} in Figure~\ref{f:ddp_vali_parameter} as square markers for later comparison with the validation data in Section~\ref{ss:resu-veri}.
In this work, we run all these models to test their performances because behind the training sets are GALAH and APOGEE spectra, which have different wavelength regions. The use of all these models will build a complete set of elements and also test the non-negligible systematics among the high-resolution surveys' labels. In addition, for some overlapped elements, we also compare their bias and dispersion, which can provide a flexible choice of models for the users in the future.
Based on results from Figure 2 in X19, Li, Sc, V, Zn, Y, Eu were removed due to the weak correlations of the gradients, and will not be considered in this study.

The process of applying the \ddpayne{} to MUSE spectra is as follows: After loading the \ddpayne{} models, same as X19, several wavelength regions are masked including areas with telluric bands, the very red wavelength region ($>8750\Angstrom$), and a region overlapped by the dichroic $(5720-6060\Angstrom)$ where the LAMOST spectral performance is low \citep{2015MNRAS.448..822X}. Each MUSE spectra is then fitted with both the \ddpayneg{} and \ddpaynea{} model. 
Following the procedure outlined in X19, we also run an additional model for stars being identified as metal-poor in \ddpaynea{} ($\feh{}<-0.6$) to improve the precision for metal-poor stars. 
This \ddpayne{} APOGEE metal-poor model (\ddpayneamp{}) is trained by X19 based on common metal-poor stars in APOGEE-\emph{Payne} and LAMOST DR5. For stars with $\feh{}<−1.0$ dex, only labels from \ddpayneamp{} model are adopted.
For stars with metallicity in $−1.0<\feh{}<−0.6$ dex, we take the weighted mean value of results from \ddpaynea{} and \ddpayneamp{} models, with weighting given below by Equation~\ref{apogeemetal}. For stars with $\feh{}>-0.6$ dex, we adopt the the \ddpaynea{} measurements.
\begin{equation}
\begin{split}
&\omega_{\rm MP} = (\feh_{\rm MR} + 0.6) / (-0.4), \\
&\omega_{\rm MR} = 1 - \omega_{\rm MP},  \\
&\text{[X/Fe]} = \omega_{\rm MR} \times \text{[X/Fe]}_{\rm MR} + \omega_{\rm MP} \times \text{[X/Fe]}_{\rm MP}.
\label{apogeemetal}
\end{split}
\end{equation}
Here MR is the label from \ddpaynea{}, and MP is the label from \ddpayneamp{}, $\omega$ is the relative weight for each label [X/Fe]. In the following sections, when we refer to the labels derived from \ddpaynea{} we mean this combined set of results.

In addition, we determine the radial velocity and its error using the Doppler equation at the same time during the fit, which is the same as \thepayne{} \citep{2019ApJ...879...69T}.

We also determine two values to represent the quality of fitting. One is the reduced $\chi^2$, calculated given by
\begin{equation}
\chi^{2}= \frac{1}{n} \sum_{i=1}^{n} \frac{\left(P_{i}-O_{i}\right)^{2}}{e_{i}^2},
\end{equation}
where $n$ is the number of wavelength pixels, $P_{i}$,  $O_{i}$ and $e_{i}$ are the best-fit spectra, observed spectra and the flux error in the $i$-th wavelength pixel.
The other is the Pearson correlation coefficient between the pixel values of the observed and the best-fit \ddpayne{} spectra.

Figure \ref{f:muse_ddp_comparison} illustrates the comparison of observed MUSE spectra (in black) with the best fit by \ddpayne{} (in red) for the same star as that in Figure \ref{f:lm_muse_comparison}. The residual spectrum is shown in the bottom panel.
The difference is less than 0.03 over most of the spectrum, including areas with strong absorption lines. This demonstrates that \ddpayne{} training models can reproduce the MUSE spectra with high fidelity.

\subsection{Assessing the precision of labels}

Even though our best-fit model spectra are quite similar to observed MUSE spectra, it is not guaranteed that all stellar labels from \ddpayne{} are estimated precisely.
Many spectral features suffer from blending due to the low spectral resolution. Since \ddpayne{} is developed to avoid astrophysical correlations, the imprecise stellar labels are due to weak or no absorption features in the fitting wavelength window.
When compared to LAMOST spectra, MUSE loses information in the blue regions $3700-4750\Angstrom$, which contains absorption features of many elements.
Therefore, we cannot expect the performance of \ddpayne{} on the MUSE spectra to be as good as that on the LAMOST spectra, even if MUSE spectra have higher SNR. To check the precision of estimated stellar labels, we follow T17 \& X19's work, and for each stellar label, we compare the gradient spectra of \ddpayne{} with Kurucz spectral model \citep{1970SAOSR.309.....K, 1993KurCD..18.....K, 2005MSAIS...8...14K} and calculate their correlation coefficient.
If a label is physically measured from the spectral absorption features, the gradient spectra predicted by \ddpayne{} should be very similar to the Kurucz spectral model, with a correlation coefficient close to 1.
Otherwise, if a weak or no absorption feature is identified for the label the correlation coefficient will be very low.
Due to the computational expense of calculating the theoretical model, we use a template of Kurucz model gradient spectra from X19 (see Table 1) which is calculated based on a list of reference stars covering a wide range
in the parameter space, from −2.5 to 0.5 dex in \feh{} and 4000
to 7000 K in \teff{}. Therefore, for each star, we compare the gradient spectra from \ddpayne{} with one of the template spectra having the closest parameter distance to the target star. The parameter distance is defined by X19 as
\begin{equation}
D=\sqrt{\left(\Delta \teff / 100 \mathrm{K}\right)^{2}+(\Delta \log g / 0.2)^{2}+(\Delta \feh / 0.1)^{2}}.
\end{equation}
Finally, the correlation coefficient of each label will be provided in the final catalog as well as a flag based on the correlation values to represent whether the label is precise or not. Here we assign the flag of 1 if the correlation coefficient is larger than 0.5, otherwise, the flag is set to 0. 
In this way, we will know which elements are determined precisely for each star.

\subsection{Building the catalog}
\label{ss:meth-catalog}

\begin{table*}
\caption{Descriptions of all the column names and information for the DD-Payne stellar label catalog.}
\label{tab:tab2}
    \begin{tabular}{lll}
    \hline
    Column & Desceiption & Unit\\
    \hline
    stellar$_{-}$id & Stellar ID of stars & String\\
    ra & Right ascension from the input catalog (or RAJ2000)  & deg\\
    dec & Declination from the input catalog (or DEJ2000) & deg\\
    snr$_{-}$muse & Spectra signal-to-noise ratio calculated by median of all pixels & \perpixel{}\\
    \teff & Effective temperature & K\\
    \teff$_{-}$err & Uncertainty in \teff & K\\
    \teff$_{-}$flag & Quality for the value based on the examination of the DD-Payne gradient spectra $\partial f_{\mathrm{\lambda}} / \partial T_{\mathrm{eff}}$ \\
    \teff$_{-}$gradcorr & Correlation coefficients of $\partial f_{\mathrm{\lambda}} / \partial T_{\mathrm{eff}}$ between the DD–Payne and the Kurucz model \\
    \logg & Surface gravity & in cgs \\
    \logg$_{-}$err & Uncertainty in \logg & in cgs \\
    \logg$_{-}$flag & Quality for the value based on the examination of the DD-Payne gradient spectra $\partial f_{\mathrm{\lambda}} / \partial \logg$ \\
    \logg$_{-}$gradcorr & Correlation coefficients of $\partial f_{\mathrm{\lambda}} / \partial \logg$ between the DD–Payne and the Kurucz model \\
    V$_\mathrm{mic}$ & Micro-turbulent velocity &  \kms \\
    V$_\mathrm{mic}$$_{-}$err & Uncertainty of micro-turbulent velocity &  \kms \\
    $V_\mathrm{mic}$$_{-}$flag & Quality for the value based on the examination of the DD-Payne gradient spectra $\partial f_{\mathrm{\lambda}} / \partial V_\mathrm{mic}$ \\
    $V_\mathrm{mic}$$_{-}$gradcorr & Correlation coefficients of $\partial f_{\mathrm{\lambda}} / \partial V_\mathrm{mic}$ between the DD–Payne and the Kurucz model \\
    \feh & Metallicity &  \dex \\
    \feh$_{-}$err & Uncertainty of metallicity &  \dex \\
    \feh$_{-}$flag & Quality for the value based on the examination of the DD-Payne gradient spectra $\partial f_{\mathrm{\lambda}} / \partial \feh$ \\
    \feh$_{-}$gradcorr & Correlation coefficients of $\partial f_{\mathrm{\lambda}} / \partial \feh$ between the DD–Payne and the Kurucz model \\
    \xfe & Chemical abundance fraction to iron, sorted by Mg, Si, Ca, Ti &  \dex \\
    \xfe$_{-}$err & Uncertainty of chemical abundance &  \dex \\
    \xfe$_{-}$flag & Quality for the value based on the examination of the DD-Payne gradient spectra $\partial f_{\mathrm{\lambda}} / \partial \mathrm{[X/H]}$ \\
    \xfe$_{-}$gradcorr & Correlation coefficients of $\partial f_{\mathrm{\lambda}} / \partial \mathrm{[X/H]}$ between the DD–Payne and the Kurucz model \\
    red$_{-}$chi2 & Reduced $\chi^2$ of the spectral fit (hereafter $\chi^{2}$)\\
    corr$_{-}$flux & Correlation coefficient of the muse and best-fitting model spectra \\
    \hline
    \end{tabular}
\end{table*}

After deriving two series of stellar labels for each star by applying \ddpayneg{} and \ddpaynea{} models and assessing the precision of each label value, we compile  two catalogs. 
Table \ref{tab:tab2} presents a description of the columns in the catalogs. Due to different science goals, wavelength coverage, and data quality, the relative performances of the two models are different for different stellar labels.
Some elements are estimated better in one model as compared to the other.
In normal circumstances, the version of the label with a larger gradient spectra correlation coefficient should be selected.
In the following Section \ref{ss:resu-veri}, we discuss in detail the bias and precision of stellar labels from these training models.

\section{Results}
\label{s:resu}

In Section \ref{s:meth} we gave detailed steps for applying the \ddpayne{} on MUSE spectra, measuring their stellar labels and building the \ddpaynea{} and \ddpayneg{} catalogs. 
To test whether our method can measure stellar labels from MUSE observations with high precision, we now do two validations: First, we compare the \ddpayne-estimated stellar labels from common stars observed by both MUSE and LAMOST; Second, we use our method to analyze the \feh{} and \mgfe{} distributions of stars in the bulge and compare our results with that of the previous high-resolution spectroscopic studies. The motivation for each validation has been discussed in detail in Section~\ref{s:data-vbd}.

\subsection{Validation 1: LAMOST-MUSE cross-validation samples}
\label{ss:resu-veri}

\begin{figure*}
\includegraphics[width=1.88\columnwidth]{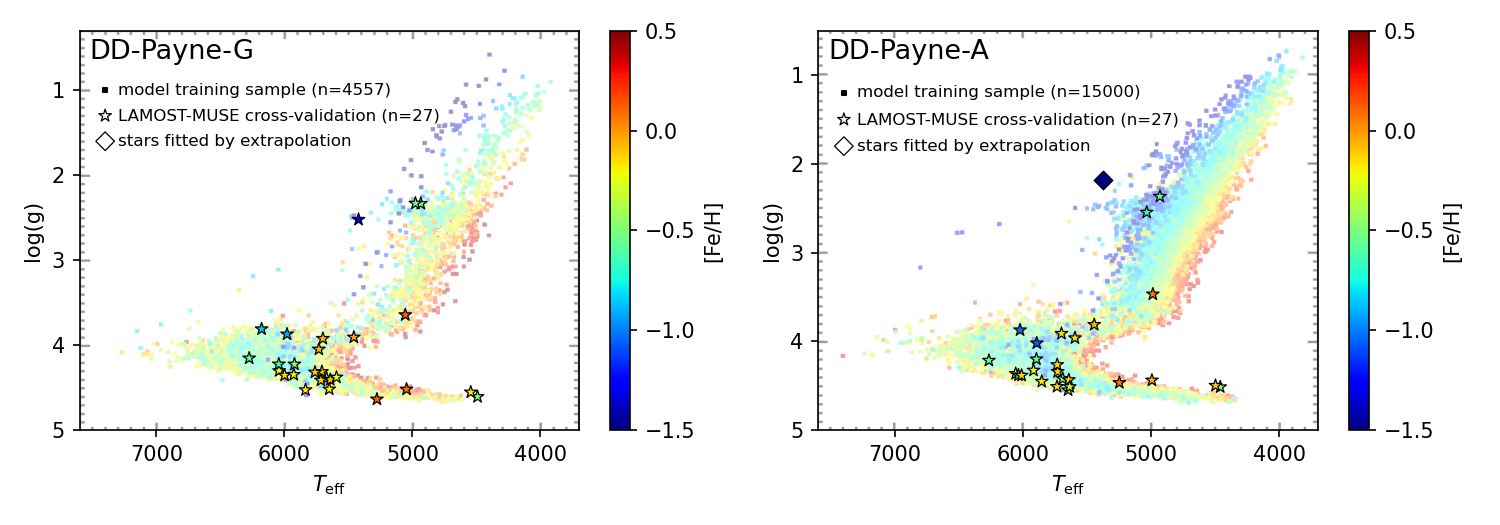}  
\caption{
Distributions of the \ddpayne{} training set (Figure 5 in X19, in square markers) and LAMOST-MUSE cross-validation samples (in star markers) in the \teff{}-\logg{} plane color-coded by \feh{}. Left panel:  Distribution of the \ddpayneg{} training stars, the \teff{} and \logg{} are from a corrected version GALAH DR2 values (see Section 3 and Appendix A of X19) The number of stars for each sample is noted in each subplot. The LAMOST-MUSE cross-validation samples are selected following the criteria of Section~\ref{ss:resu-veri} and span reasonable parameter ranges than those from  \ddpayne{} training sets. No stellar parameter is measured by extrapolation.
}
\label{f:ddp_vali_parameter}
\end{figure*}

We follow the steps described in Section \ref{s:meth} and derive stellar labels of common stars in LAMOST and MUSE using both \ddpayneg{} and \ddpaynea{} models. 
To ensure the precision of labels, we use the selection criteria given by
\begin{equation}
\left\{\begin{array}{l}
\snrlm{}>35 \text{pix}^{-1} (\sim \rm{SNR}_{\rm{LAMOST}}^{\rm{g-band}}>23 \text{pix}^{-1}) \\
\snrmuse{}>80 \text{pix}^{-1} \\
\end{array}\right.
\label{eqn:thresholdval1}
\end{equation}
Here \snrlm{} and \snrmuse{} is the median signal-to-noise ratio of the LAMOST and MUSE spectra, respectively.
We made a cut-off of \snrlm{} to ensure the correction of stellar parameters measured from LAMOST spectra, this threshold is looser than the criteria of \ddpayne{} training set ($\rm{SNR}_{\rm{LAMOST}}^{\rm{g-band}}>30$ \perpixel{}) to keep more common stars. The threshold of \snrmuse{} cut-off is more strict than that of \snrlm{} because the SNR of MUSE spectra is somehow overestimated after data reduction pipeline, which is confirmed by looking at the spectra of the two validation data and the distribution of flux residuals of the fitting divided by the flux error (the right panel of Figure~\ref{f:muse_ddp_comparison}).
Nevertheless, most MUSE spectra have SNR \perpixel{} greater than 200.
This is because these observations typically have long exposure time, as they were targeting external galaxies which are much fainter than the stars we are interested in.
Finally, we end up with 27 common stars in \ddpayneg{} and \ddpaynea{} results, during which the $\chi^2$ is mostly concentrated in 4$\sim$25, meaning the \snrmuse{} is overestimated by a factor of 2$\sim$5.
Star markers/diamonds in Figure~\ref{f:ddp_vali_parameter} represent this cross-validation sample. These stars span a reasonable range of \teff{}, \logg{} and \feh{} than those from \ddpayne{} training sets. One star is measured by extrapolation, which will be discussed in Section~\ref{sss:resu_veri_some}.

\subsubsection{Bias and precision check for main labels}
\label{sss:resu_veri_some}

\begin{figure*}
\includegraphics[width=1.53\columnwidth]{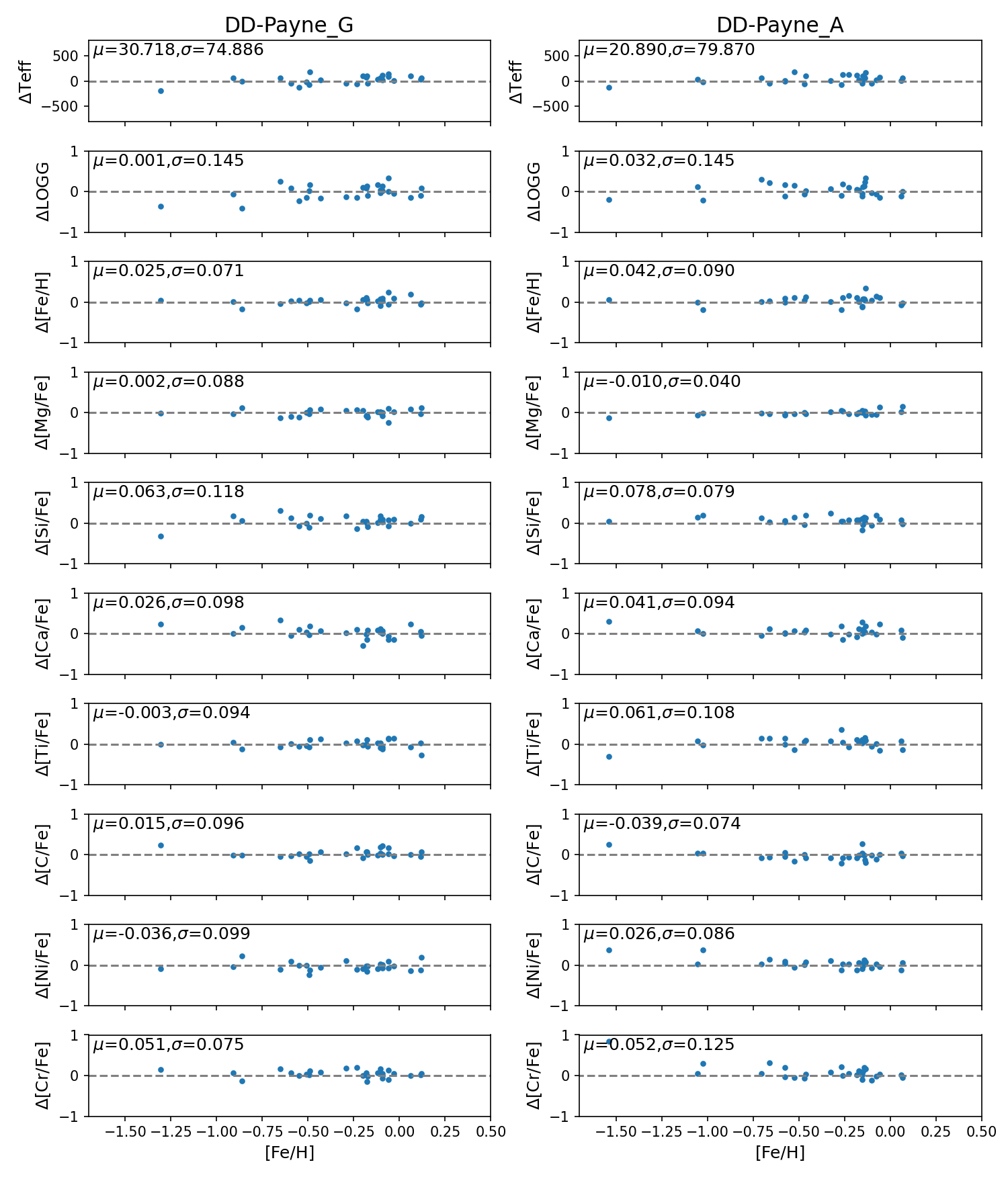}
\caption{
Differences of stellar labels derived from MUSE spectra with those from LAMOST spectra using \ddpayneg{} (left column) and \ddpaynea{} (right column) models as a function of \feh from MUSE spectra, respectively. All the stars in these plots have LAMOST spectra with SNR higher than 35 \perpixel{}. The median and dispersion of all points are marked in the upper left corner of each panel, the dispersion is calculated as half of the difference between 15.87 and 84.13 percentile values. 
}
\label{f:com_ddp}
\end{figure*}

In this section, we focus on \teff{}, \logg{} and several chemical abundances which we reported with confidence in the abstract.
Figure \ref{f:com_ddp} shows the differences of \ddpayne-estimated stellar labels from MUSE and LAMOST spectra, as a function of \feh{}.
The median and dispersion of all points are marked in the upper left corner of each panel, where the dispersion is calculated as half of the difference between 15.87 and 84.13 percentile values. The median and dispersion are indicative of the bias {$\mu$} and precision {$\sigma$} of our label estimates, respectively.
The left and right panels show results for the \ddpayneg{} and \ddpaynea{} models, respectively. 
This figure demonstrates that the dispersion of $\Delta$\teff{} is about 75K, $\Delta$\logg{} is around 0.15 dex for both the \ddpayneg{} and \ddpaynea{}. As for chemical abundances, the dispersion is less than 0.1 dex in general and we consider this to be precise, as this is roughly the uncertainty in the LAMOST labels themselves.
Therefore, most of the stellar labels from MUSE spectra and LAMOST spectra are in good agreement in Figure~\ref{f:com_ddp}. 
Even though the dispersion of some elements (e.g. \sife{} from \ddpayneg{} and \crfe{} from \ddpaynea{}) is slightly higher, it is still close to 0.1 dex. 
The $\alpha$-elements listed in this figure are some of the most useful ones as they can be used to constrain intrinsic stellar properties such as evolutionary stage, distance, birth radius, extinction, and even age \citep{2019ApJ...883..177N, 2020arXiv201113745H, 2022MNRAS.510..734S}.
Therefore, this agreement shows the usefulness of \ddpayne{} based estimates from MUSE observations for Galactic studies.
The bias in this figure reflects the differences between LAMOST and MUSE instruments and the different performances when losing wavelength in $3700-4750\Angstrom$. However, it can be seen that all of them are smaller than the dispersion. Therefore, in this work we do not perform the bias correction on any label. Whether to do it will be re-investigated when we have a larger sample.

In addition, by comparing the dispersion of the same elements under different model estimates, we also found that except \feh{}, \tife{} and \crfe{}, all the labels calculated using the \ddpaynea{} model have lower dispersion than that from the \ddpayneg{} model.
The dispersion of \mgfe{} from the \ddpaynea{} model is even less than 0.05 dex.
Since we derived labels from the same spectra, the reason must be the relative performance of the different models. Additionally, when comparing the derived gradient spectra with Kurucz spectral model for these labels, \ddpayneg{} with lower correlation coefficients, which indicates \ddpaynea{} model has better performance. This is because the training set of \ddpaynea{} is larger than that of \ddpayneg{}, so the measured stellar labels will be more precise.


There is one star in the panel of \ddpaynea{} measured by extrapolation since the \feh{} is less than -1.5 dex, which is shown in the right panel of Figure 4 as a "diamond". By looking at the parameter differences in each panel, we can see from Figure 5 that \teff{}, \logg{}, \feh{}, \mgfe{} and \sife{} measured from MUSE spectra of this star have good agreements with those measured from LAMOST spectra. But for the other parameters, the differences are larger. Note that the LAMOST spectra of this star was also measured by extrapolation, whether this indicates anything about extrapolation will be investigated when we have more stars.

\subsubsection{Bias and precision check for all labels}
\label{sss:resu-accuracy}

\begin{figure*}
\includegraphics[width=1.75\columnwidth]{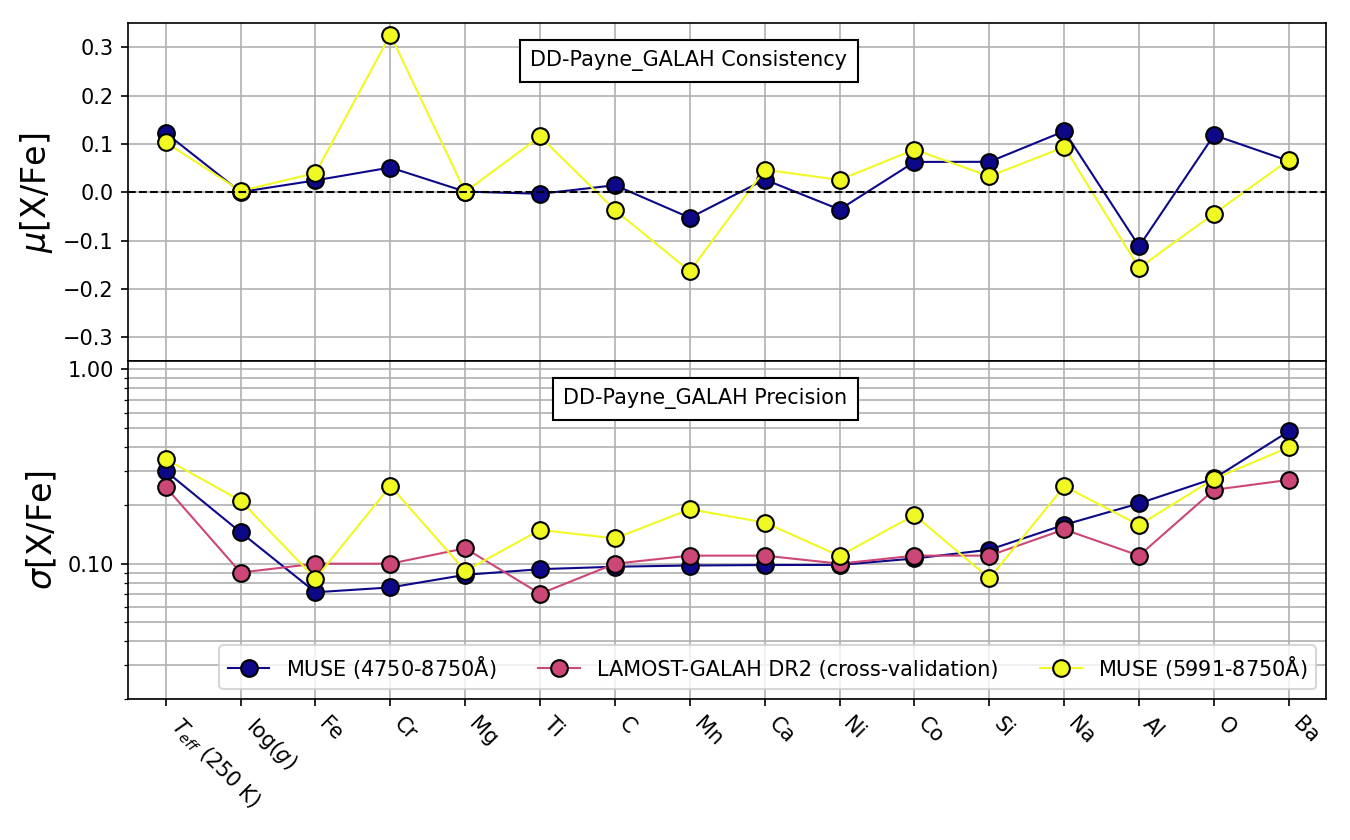}
\caption{
The bias (top panel) and precision (bottom panel) of all stellar labels derived from \ddpayneg{} model, respectively. The bias and precision are calculated in the same way as Figure \ref{f:com_ddp} by taking the median and half the difference between 15.87 and 84.13 percentile values. We list all measured labels and arrange them from small to large according to the precision of the elements. We also plotted results in the Bulge case which loses pixels between $(4750-5991\Angstrom)$. From these two plots, for the \ddpayneg{} model, Fe, Mg, Ni, Si are precisely estimated with $\sigma$ close to or better than 0.1 dex when fitting MUSE spectra between both $(4750-8750\Angstrom)$ and $(5991-8750\Angstrom)$. 
As for bias, most of the elements are within 0.1 dex except Co, Mn, Na, and Al.
}
\label{f:acc_galah}
\end{figure*}

\begin{figure*}
\includegraphics[width=1.75\columnwidth]{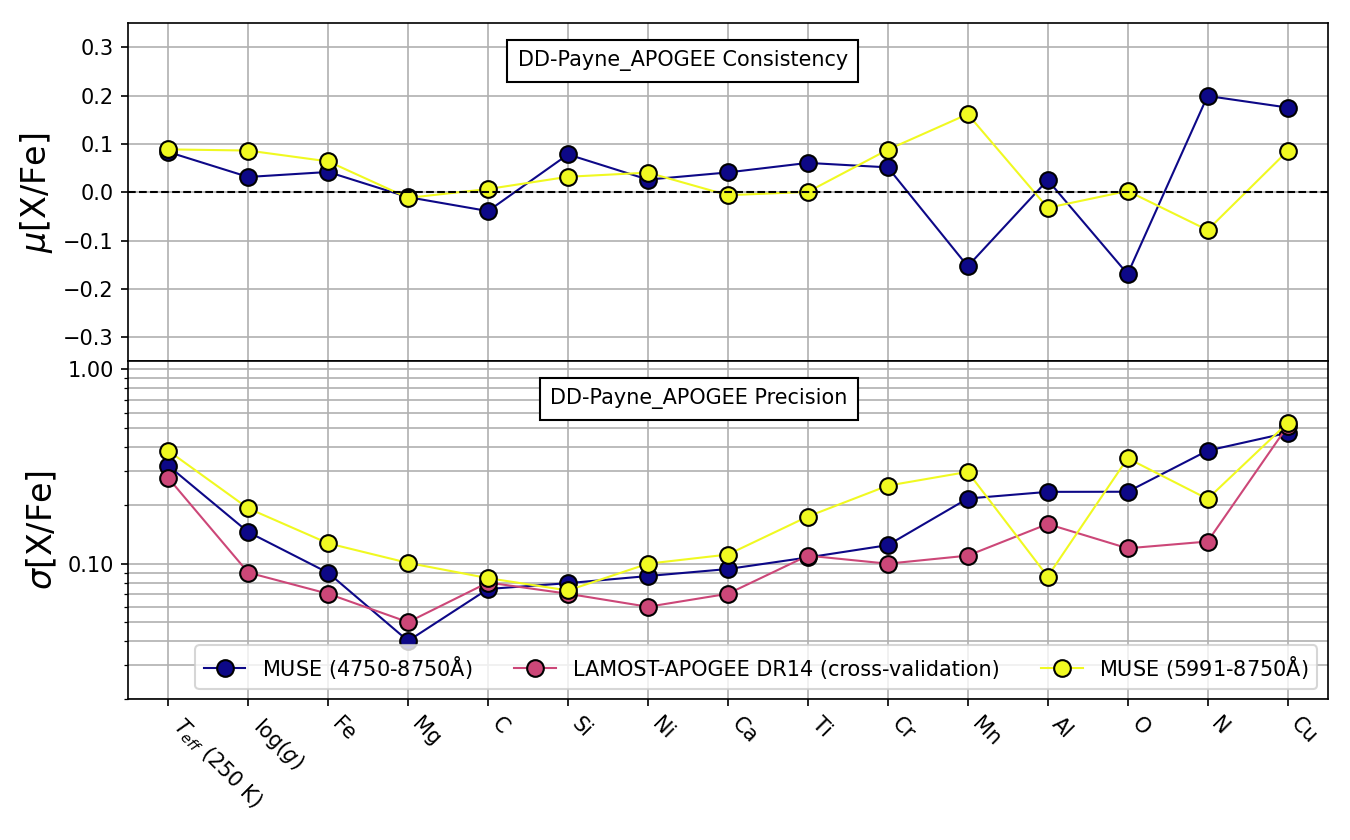}
\caption{
The same plot as Figure \ref{f:acc_galah} but now using \ddpaynea{} model. From these two plots, for \ddpaynea{} model, Fe, C, Si, Mg, Ni, and Ca are well estimated with $\sigma$ less than 0.1 dex when fitting MUSE spectra between both $(4750-8750\Angstrom)$ and $(5991-8750\Angstrom)$. Besides, MUSE $(5991-8750\Angstrom)$ shows a larger bias and dispersion. As for bias, most of the elements are within 0.1 dex except Mn, O, N, and Cu. Comparing with results from \ddpayneg{} model, for the same element, \ddpaynea{} model has better precision.}
\label{f:acc_apogee}
\end{figure*}

Other than the parameters discussed in the previous section, we also estimated abundances for additional elements, with 23 stellar labels for \ddpayneg{} and 16 labels for \ddpaynea{}. However, the dispersion for many of the abundances is not shown as they are much larger than those analyzed in Figure~\ref{f:com_ddp}. 
This is because, unlike LAMOST, MUSE does not have any pixels in the wavelength range $3700-4750\Angstrom$, which contains a large number of absorption features. 
In this section, we will further discuss the precision of these labels and compare them with the results of cross-validation samples between LAMOST and GALAH or APOGEE.

We perform the same analysis in Figure~\ref{f:com_ddp}, and derive bias ($\mu$) and the dispersion ($\sigma$) for all labels in Figure \ref{f:acc_galah} and Figure \ref{f:acc_apogee}.
In each plot stellar labels from MUSE are determined in two regions: $\lambda=(4750-8750\Angstrom)$ (in blue) and $\lambda=(5991-8750\Angstrom)$ (in yellow), corresponding to normal spectra and spectra in high extinction fields such as the Galactic bulge. 
We also over-plot the dispersion of label differences between those from LAMOST spectra and GALAH DR2 \citep{2018MNRAS.478.4513B} or APOGEE-\emph{Payne} \citep{2019ApJ...879...69T} in red.

For the LAMOST-MUSE cross-validation sample (the blue line), elements such as Na, Al, O, and Ba in \ddpayneg{} model and Mn, Al, O, N, and Cu in \ddpaynea{} have high median and dispersion (larger than 0.1 dex).
This can be due to two reasons.
We find that the bias and dispersion of Na, Al, O, and Ba in cross-validation samples are also high. 
For these elements, the high dispersion is the inheritance from the \ddpayne{} models. 
However, for Mn and N, the high dispersion is due to the loss of the blue wavelength region ($3700-4750\Angstrom$) from the MUSE spectra compared to the larger wavelength region covered by LAMOST.
This can be seen in Figure~\ref{f:acc_apogee}, where $\mu$ and $\sigma$ show a large discrepancy between the LAMOST-GALAH/APOGEE cross-validation estimates and the LAMOST-MUSE cross-validation estimates.
This is further confirmed by the gradient spectra of these two elements (See Figure 22 in X19), where the dominant absorption features of Mn and N are in $3700-4750\Angstrom$, which the MUSE spectra are lacking.

In general, for most of the labels in these two figures, the dispersion of the cross-validation estimates and MUSE estimates are in agreement.
Therefore, it can be verified that the models designed for LAMOST spectra can be applied to MUSE spectra, and measure some of labels (\teff{}, \logg{}, \feh{}, [Ni/Fe], [Cr/Fe], [C/Fe] and $\alpha$ abundances) with similar precision.
It is worth noting that there are several elements in these figures in which MUSE has higher precision than the LAMOST cross-validations, for example, Fe, Cr, Mg in \ddpayneg{} and Mg in \ddpaynea{}, this is due to the small number of stars in the common sample of LAMOST and MUSE. 

For the LAMOST-MUSE cross-validation estimates in the wavelength region of $\lambda=(5991-8750\Angstrom)$ (the yellow line). Both bias and dispersion are larger than when it is fitted in $4750-8750\Angstrom$ (the blue line).
This is because more absorption features in the region of $4750-5991\Angstrom$ are lost.
Nevertheless, some elements such as \feh{}, \mgfe{}, \sife{} in \ddpayneg{} and \feh{}, \mgfe{}, \cafe{}, \cfe{}, \nife{}, \sife{} in \ddpaynea{} still have dispersion lower than 0.1 dex.
This indicates that after further losing pixels in the wavelength range of $4750-5991 \Angstrom$ due to high extinction from the bulge, we still can recover some stellar labels for these fields.

\subsubsection{Theoretical precision prediction}
\label{sss:resu-veri-internal}

\begin{figure*}
\includegraphics[width=1.64\columnwidth]{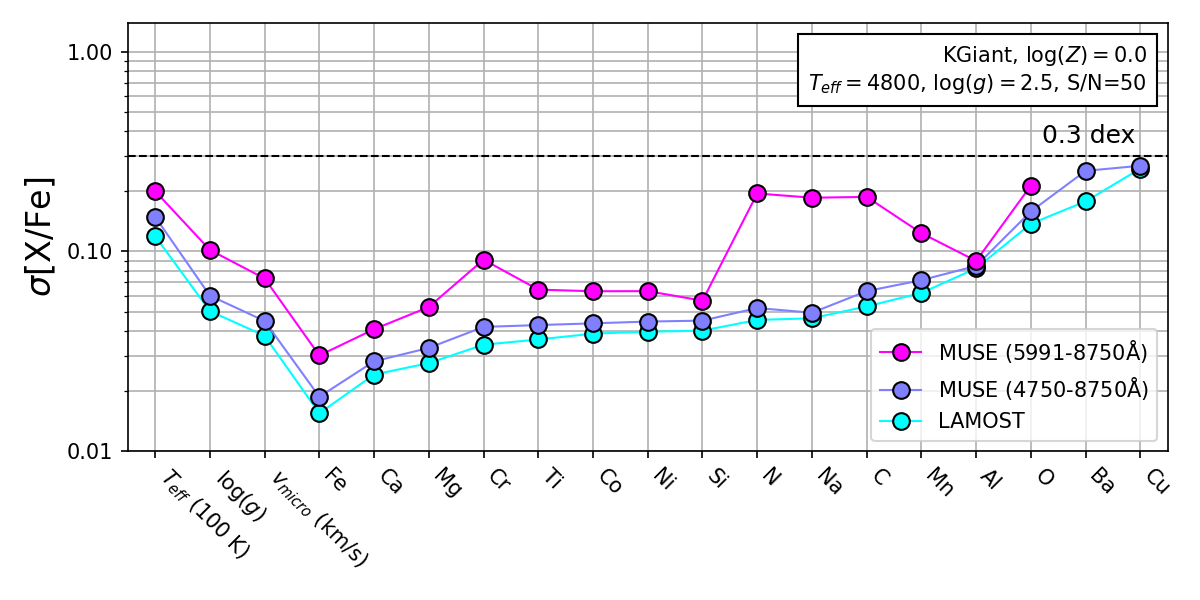}
\caption{
Estimation of the theoretical precision for a typical solar-metallicity K-Giant with a g-band SNR of 50 \perpixel{} in three different scenarios. The LAMOST spectra cover the wavelength range between $3700-8750 \Angstrom$. Both MUSE $(4750-8750 \Angstrom)$ and MUSE $(5991-8750 \Angstrom)$ have the same resolution as LAMOST. The theoretical precision is predicted by using the \crlb{} method which is an efficient method for computing the expected precision of stellar labels determined via full spectral fitting \citep{2020ApJS..249...24S}. When losing wavelength pixels in wavelength $3700-4750 \Angstrom$ (LAMOST versus MUSE $(4750-8750 \Angstrom)$, the precision of all the labels is getting slightly worse by an average factor of 1.17, but there is no label having a large change. When further losing wavelength pixels in the range of $4750-5991\Angstrom$ (i.e., for MUSE cubes in high extinction fields), the precision of all the labels is worse by an average factor of 2.16. However, there are some labels such as Cr, N, Na, and C which have much larger shifts towards larger uncertainties by a factor of 2.66, 4.30, 4.02, 3.53, respectively due to the loss in wavelength coverage.}
\label{f:inter_preci}
\end{figure*}

In Section~\ref{sss:resu_veri_some} and Section~\ref{sss:resu-accuracy} we discussed the bias and precision of stellar labels from MUSE spectra in comparison with results from LAMOST spectra and discussed the influence of losing wavelength pixels in $3700-4750\Angstrom$ and $4750-5991\Angstrom$.
The theoretical precision, which is the standard deviation of each label X (written as $\mathrm{X_{-}err}$ in Table~\ref{tab:tab2}) as a by-product of \ddpayne{},  will also be affected by the changes in wavelength coverage.
Since we don't have enough LAMOST and MUSE common stars with a given SNR, we use the method \crlb{} \citep{2020ApJS..249...24S} to forecast the theoretical precision, which uses \abinitio{} spectral model and the Cramér-Rao Lower Bound (CRLB)
to quantify the chemical information content of stellar spectra in terms of theoretical precision.
Figure~\ref{f:inter_preci} shows the predicted theoretical precision of stellar labels for a typical solar-metallicity K-Giant with a g-band SNR of 50 \perpixel{} in three different scenarios.
By comparing these scenarios, we find that when losing wavelength range in $3700-4750\Angstrom$ (cyan versus blue), the theoretical precision of all the labels is getting slightly worse by an average factor of 1.17, but there is no label having a large change.
When further losing wavelength pixels in wavelength $4750-5991\Angstrom$ (blue versus pink), the precision of all the labels is getting worse by an average factor of 2.16.
However, there are some labels such as Cr, N, Na, and C which have a much greater theoretical precision change with a factor of 2.66, 4.30, 4.02, 3.53, respectively, as we lose critical absorption lines in a more restricted wavelength range.
This is shown in the gradient spectra of these labels from Figure 22 of X19, where they show that there are no significant features for these abundances in the remaining wavelength range.
Fortunately, precision of labels like \teff{}, \logg{}, \feh{}, and some $\alpha$ elements are not significantly affected.

\subsection{Validation 2: observations towards the Galactic bulge}
\label{ss:resu-bulge}

From the above sections we have shown that we can derive stellar labels precisely for individual stars in MUSE observations via the utilization of the \ddpayne{}. Here we want to demonstrate that our method is also useful to determine stellar labels for stars in dense fields. The Milky Way bulge is one of the densest regions of the Galaxy. 
In the past decades, there have been many fiber-fed spectroscopic surveys focusing on the bulge to study its chemical distribution such as Gaia-ESO \citep{2012Msngr.147...25G}, ARGOS \citep{2013MNRAS.430..836N}, and GIBS survey \citep{2002Msngr.110....1P}, which have revealed multiple stellar components with different mean metallicity. 
A highlight of this study is from \citealt{2019A&A...626A..16R} who revealed the bimodality in \mgfe{}-\feh{} distribution using the bulge stars from APOGEE DR14 \citep{2018ApJS..235...42A}.
In terms of the integral-field spectroscopic survey, \citealt{2018A&A...616A..83V} observed four fields with MUSE towards the inner bulge and measure radial velocity based on the CaT region of spectra and reached $\sigma V_{\mathrm{gc}}\sim$ 140 \kms. They also found that the velocity dispersion peak is symmetric with respect to the distance $z$ from the Galactic plane.
To summarize previous bulge studies, chemical element distributions have been studied based on high-resolution fiber-fed spectroscopic surveys, and kinematics has been studied using IFS data.
Since our method can precisely determine stellar labels from MUSE spectra, this should enable us to study the chemodynamics of the Milky Way bulge using IFS data.

\subsubsection{Data refinement}
\label{sss:resu-bulge-data}

\begin{figure*}
\includegraphics[width=1.89\columnwidth]{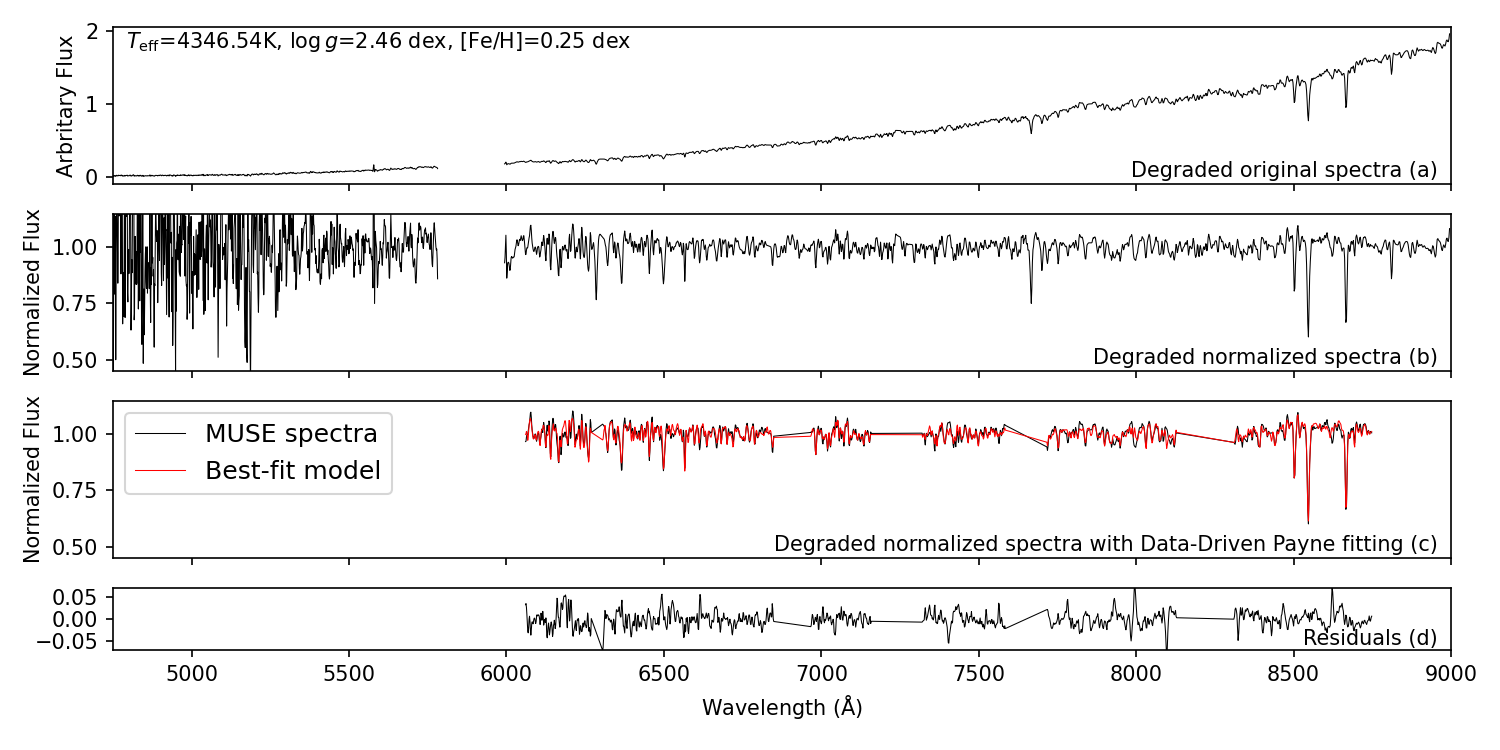}
\caption{
The detailed process of spectral processing of one bulge star in the MUSE data-cube. (a) Original MUSE spectra after degrading which demonstrates the high extinction in the Galactic center causing almost no flux in short wavelength. (b) Spectra after normalization, which shows the extremely high extinction makes the short-wavelength spectra between $(4750-5991\Angstrom)$ impossible to fit. (c) The predicted spectra and original MUSE spectra of this star. (d) The residuals of predicted spectra and the original MUSE spectra.
}
\label{f:bulge_emp}
\end{figure*}

\begin{figure*}
\includegraphics[width=1.89\columnwidth]{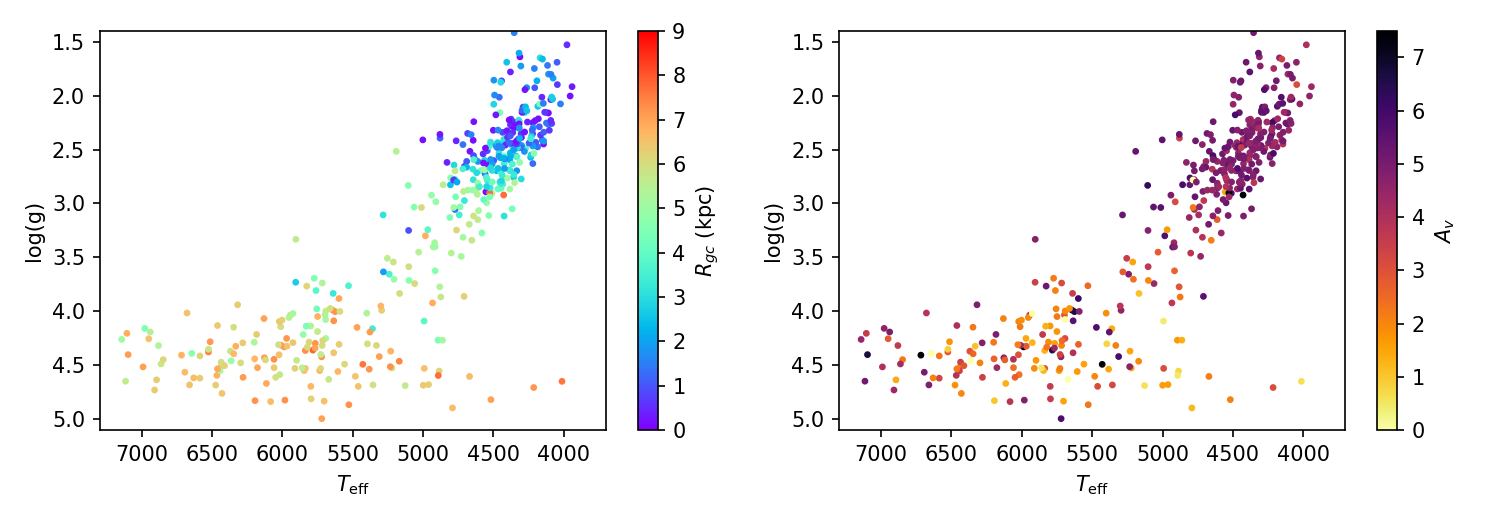}
\caption{
H-R diagrams of stars with good fitting to the model defined by Equation~\ref{eqn:thresholdval21}. The left-hand side panel is color-coded with \rgc{} (distance to the Galactic center) in cylindrical Galactocentric coordinates, where distance from the Sun to the Galactic Center is adopted as \Rsun{}$=8.2$ kpc, $z_{\odot}=25$ pc \citep{2016ARA&A..54..529B}. The right-hand side panel is color-coded with extinction $A_V$. Both \rgc{} and $A_V$ are predicted by BSTEP \citep{2018MNRAS.473.2004S} taking parameters of \teff{}, \logg{}, \feh{}, \mgfe{} from DD-Payne and $m_{J}$, $m_{Ks}$, $m_{H}$ magnitude from VVV DR2.
}
\label{f:hrbulge}
\end{figure*}

The bulge data processing begins with the steps outlined in Section~\ref{s:meth}.
The spectra are extracted from the 29 bulge data-cubes using \pampelmuse{} with VVV DR2 \citep{2017yCat.2348....0M} as the input photometry catalog. Next, the spectra are degraded to LAMOST spectral resolution of R$\sim$1800. 
Figure \ref{f:bulge_emp} shows the degraded spectra, both raw and normalized, of one bulge star (VVV{-}J175141.74{-}282207.85) extracted from the data-cube. 

The bulge stars pose several challenges for data reduction. 
They have large heliocentric distances, which means they are typically faint and have low SNR. 
These stars also typically lie along lines of sight with high extinction, making SNR quite low in the shorter wavelength regions. 
Finally, subtracting the night sky emission lines from faint stellar spectra is challenging. 
This highlights the challenges associated with reducing the spectra in the bulge region and reinforces the need to adopt additional data refinement procedures.

To remove emission lines, we attempted to apply the Zurich Atmosphere Purge (ZAP) \citep{2016MNRAS.458.3210S}, which is an approach based on principal component analysis (PCA) to reduce emission lines for faint spectra in data-cubes. 
However, using ZAP has some undesirable side effects. 
By adjusting some tunable parameters in ZAP, we can increase the efficiency of subtracting the emission lines, but then ZAP has a significant impact on the absorption lines which are desirable for estimating chemical abundances. Strangely, the spectra of bright stars were also found to become noisy.
In the end, we find the most effective way to deal with emissions is to replace emission pixels with the average value of adjacent non-emission pixels 
In this way, we can retain more pixels in the spectra.

In addition, due to extinction in the bulge, the flux and hence the SNR in short-wavelength ranges is not enough to extract useful information (See panel (a) and (b) in Figure~\ref{f:bulge_emp}).
The NaD doublet line $(5803-5966\Angstrom)$ is also masked in data-cubes that are observed using the adaptive optics module.
Therefore, we decided to discard information for wavelengths less than $5991\Angstrom$. 
We also mask the following telluric bands; $6270-6305\Angstrom$, $6850-6966\Angstrom$, $7160-7320\Angstrom$, $7585-7715\Angstrom$, $8130-8310\Angstrom$. 
The widths of the telluric bands are narrower than those used by X19, this was done to increase the number of useful wavelength pixels. 
This range was determined after many tests to ensure that the unmasked pixels provide useful information about the chemical elements. Figure~\ref{f:acc_galah} and Figure~\ref{f:acc_apogee} can indicate that when losing pixels in $4750-5991\Angstrom$, the dispersion of some elements is still below 0.1 dex.
In panel (c) of Figure \ref{f:bulge_emp} shows the best-fit model spectra in red, alongside the original spectra in black. It is clear from the comparison between LAMOST and MUSE samples presented in Figure~\ref{f:muse_ddp_comparison}, that the spectra of bulge stars have more noise, due to a combination of these stars having higher extinction and lower average SNR.
In addition, the distribution of residuals between the observed and the best-fit model spectra for a bulge star (Figure~\ref{f:bulge_emp} panel (d)) is larger than that for the star in Figure~\ref{f:muse_ddp_comparison}, which shows the difficulty of fitting spectra in the bulge fields.

After these refinement and masking steps, stellar labels are derived from two \ddpayne{} models. We adopt the labels from the \ddpaynea{} model and elect not to use the \ddpayneg{} model to compare directly with high-resolution APOGEE data which has ample stars in the bulge. 
Given the loss of precision due to the omission of the blue wavelength region, we only adopt the \mgfe{} when studying $\alpha$ abundances for the bulge fields.

Our final MUSE bulge catalog is comprised of 8720 individual spectra containing 2721 unique stars, with the following columns 
\teff{}, \logg{}, \feh{} and \mgfe{}.
On average, from each cube about 300 spectra were extracted and 40\% of them have $\snrmuse{}>80$ \perpixel{}.
Compared with typical MUSE wavelength coverage, the bulge spectra lose the bluer regions of the spectrum, between $4750-5991\Angstrom$, so we adopt more strict selection criteria compared to our LAMOST verification sample with
\begin{equation}
\left\{\begin{array}{l}
\snrmuse{}>80 \text{pix}^{-1} \\
\rm{corr_{-}flux}>0.85 \\
\end{array}\right.
\label{eqn:thresholdval21}
\end{equation}
Here the SNR$_{\mathrm{MUSE}}$ is the SNR \perpixel{} in wavelength higher than $5991\Angstrom$. Since there are multiple spectra for each star, we selected the observation having the highest corr$_{-}$flux between the MUSE and model-predicted spectra, to represent the most physical fitting.

\begin{figure}
\includegraphics[width=0.95\columnwidth]{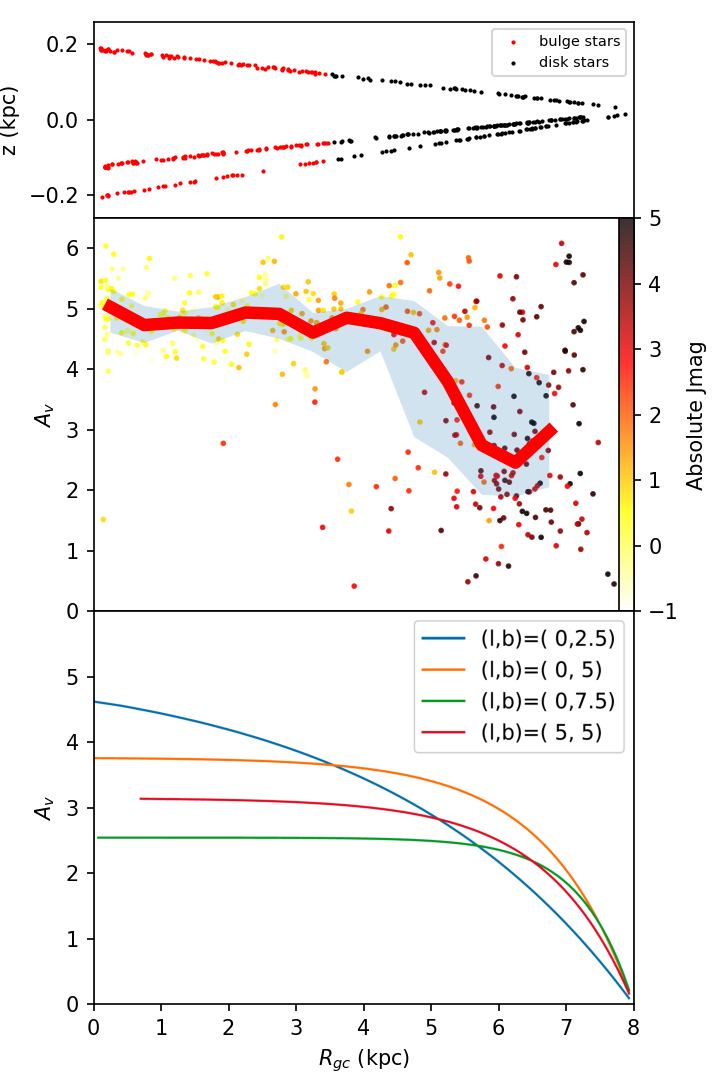}
\caption{
\textbf{Top:} Spatial distribution in cylindrical Galactocentric coordinates (\rgc{},z) of stars observed in three nine bulge fields. The distance to the Galactic center is adopted as \Rsun{}$=8.2$ kpc, $z_{\odot}=25$ pc, same as Figure~\ref{f:hrbulge}. The red dots are stars with \rgc{}$<3.5$kpc.
\textbf{Middle:} Extinction-\rgc{} distribution of all-stars with good fitting to the model defined by Equation~\ref{eqn:thresholdval21} in dots color-coded by absolute J mag. The red curve shows the median value of each distance bin and the blue fill shows the standard deviation of each bin. 
\textbf{Bottom:} Extinction $A_V$ as a function of Galactocentric radius $R$ for various lines of sight from the Sun. The extinction is based on an analytical model from \citealt{2011ApJ...730....3S} consisting of an exponential disc with warp and flare; scale length $R_{\rm d}=4.2$ kpc and scale height $0.88$ kpc. From lines towards the center, we could see that with the larger line of sight latitude, the extinction will become a constant more quickly, which is in agreement with the left bottom plot. 
}
\label{f:avrgczgal}
\end{figure}

\subsubsection{Distance estimation and its relationship with extinction}
\label{sss:resu-bulge-dist-extin}

To discriminate bulge stars from other foreground stars, we make use of stellar distance. The distance was estimated by BSTEP (Bayesian stellar labels estimator) \citep{2018MNRAS.473.2004S}, a Bayesian scheme that is used for predicting stellar ages, distances, and extinction from a set of observables given by
\begin{equation}
\mathbf{y}=\left(l, b, m_{J}, m_{K s}, T_{\mathrm{eff}}, \log g,[\mathrm{Fe/H}]_{\rm eff}\right),
\end{equation}
Briefly speaking, for each star, if we know its position, \teff{}, \logg{}, \feh{} and magnitude, its intrinsic parameters such as distance, age, and extinction can be derived by making use of isochrones \citep{2017ARA&A..55..213S}. The isochrones employed in this study are from \citealt{2012MNRAS.427..127B}. 
For our catalog, parameters of \teff{}, \logg{}, \feh{} are determined by \ddpayne{}. As for $m_J$, $m_{Ks}$, they are derived from VVV DR2 catalog.  

Figure \ref{f:hrbulge} shows the H-R diagram of stars with model spectra with good fitting to the model defined by Equation~\ref{eqn:thresholdval21} and  color-coded by Galactocentric cylindrical radial distance \rgc{} and extinction $A_V$. 
We identify bulge stars with \rgc$<3.5$kpc. 
The left panel shows that bulge stars are predominantly giants. The right panel shows that bulge stars have extinction $A_V$ in the range 4-6 mag. Note, there is a selection bias in the sense that low extinction stars are more likely to be detected.
In fact, extinction towards the Galactic center can be as high as $A_V=50$ mag \citep{2016MNRAS.456.2692N}.

In Figure~\ref{f:avrgczgal} we explore the extinction in more detail.
The top panel shows the $(R_{\mathrm{gc}} ,z)$ distribution of stars in nine MUSE cubes. The line of sight of three fields can be seen. 
Stars with distances beyond the Galactic center have low temperature ($\teff{}<4300$ K) and surface gravity ($\logg{}<1$), which fall outside of the label ranges for which we can reliably determine $\alpha$ abundances, so these stars were removed.
The middle panel of Figure~\ref{f:avrgczgal} shows a scatter plot of extinction and \rgc{} color-coded by absolute magnitude in $J$ band.
The median value of extinction is shown by the red line, along with 25 and 75 percentile dispersion as a shaded blue region.
This figure demonstrates that MUSE can successfully extract bulge giant stars with extinction up to $A_V=6$ mag.
To understand the trend of extinction with \rgc{}, in the bottom panel of Figure~\ref{f:avrgczgal} we show extinction $A_V$ as a function of Galactocentric radius \rgc{} for various lines of sight from the Sun as predicted by a theoretical model in \textit{Galaxia} \citep{2011ApJ...730....3S} consisting of an exponential disc with warp and flare; scale length $R_{\rm d}=4.2$ kpc and scale height $0.88$ kpc. 
It can be seen that if the latitude $|b|$ is greater than 5 degrees, the extinction first increases as we go towards smaller $R$, but then flattens and reaches a plateau at distances 1-2 kpc away from the sun.
A similar trend is seen in the observed data in the middle panel. This flattening is due to the finite height of the dusty disc responsible for the extinction.
For a given line of sight as long as we are in the dusty disc, the extinction increases with distance, but once we are out of the dusty disc there is no further increase in extinction.
The model qualitatively explains the trend seen in the observed data, but there are differences. The extinction for $b=5$ degrees is not high enough compared to our observations. For lower $|b|$ extinction can be high but the rise with distance is much slower. 
These differences suggest that the scale height of the disc in the model is too high.

\subsubsection{\mgfe{}-\feh{} distribution in the Galactic bulge}
\label{sss:resu-bulge-mgfe}

\begin{figure*}
\includegraphics[width=1.80\columnwidth]{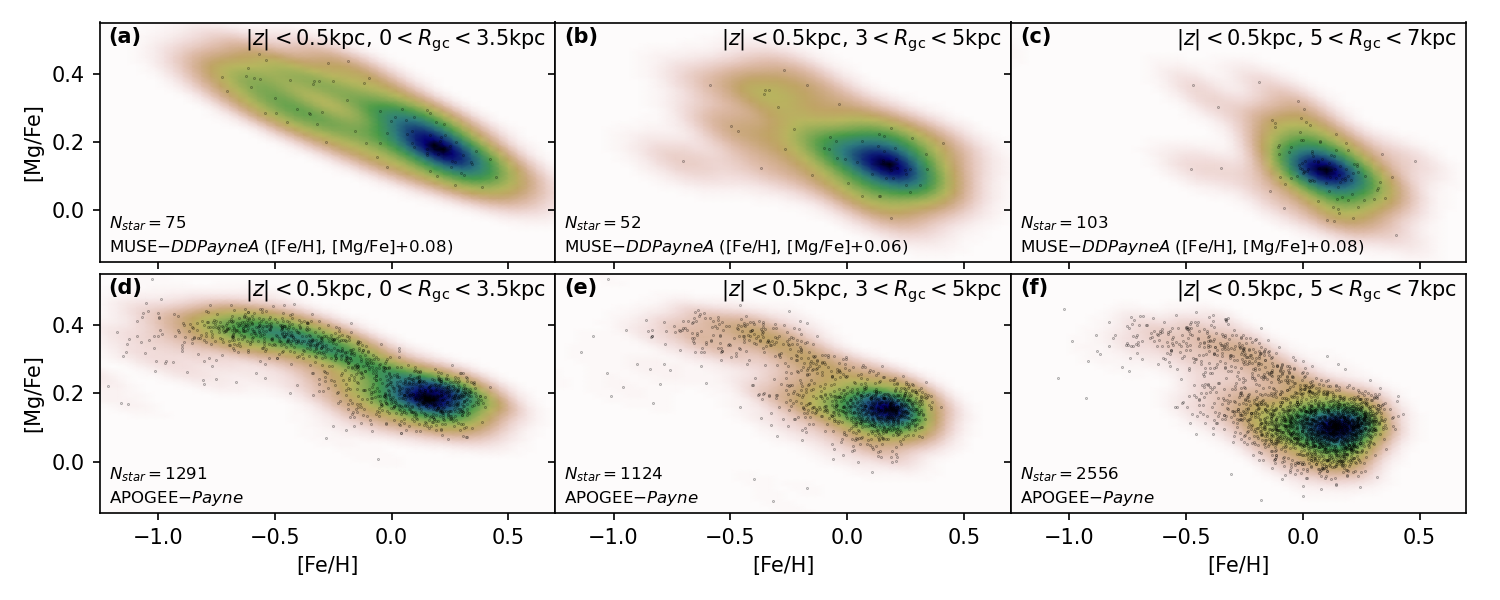}
\caption{
Distribution of stars satisfying $|z|<0.5$ kpc in the (\mgfe{}-\feh{}) plane as a function of \rgc{}, with \rgc{} increasing from left to right.
\textbf{Top}: The \mgfe{}-\feh{} distribution of stars in APOGEE-\emph{Payne} \citep{2019ApJ...879...69T}. \textbf{Bottom}: The same distribution of stars in MUSE data-cubes towards the Galactic bulge with abundances measured by \ddpaynea{} model (hereafter MUSE-\ddpaynea{}). The labels are calibrated by adding an offset for \mgfe{}, which is due the \mgfe{}-\teff{} trend inherited from APOGEE-\emph{Payne} (see details in Appendix~\ref{app:mg_teff_trend}), respectively.
In each panel, the black dots mark stars in this location and the distribution is smoothed by a Gaussian kernel function and then used to plot the color maps. 
For each \rgc{} bin, the APOGEE-\emph{Payne} and MUSE-\ddpaynea{} distributions are in agreement with the densest peak at the same location. This validates the precision of the MUSE results in dense regions.
}
\label{f:mgfe_bulge}
\end{figure*}

We next analyze the abundance distribution of stars towards the bulge. We make some additional quality selections to guarantee all the stars have precisely estimated parameters: 
\begin{equation}
\left\{\begin{array}{l}
\feh>-1.5 \mathrm{dex} \\
\teff>4300 \mathrm{K} \\
\mathrm{\mgfe{}_{-}gradcorr}>0.65 \\
\chi^2<20 \\
\end{array}\right.
\label{eqn:thresholdval22}
\end{equation}
Here stars with \feh{}$<-1.5$~dex 
and \teff{}$<4300$K are removed as it has been demonstrated in X19 that \mgfe{} for these giants are erroneous with low gradient spectra correlation coefficients (see their Figure 2).
Moreover, we apply a cut in the $\chi^2<20$ to remove stars with poor fittings from \ddpayne{} according to the $\chi^2$ distribution of the LAMOST-MUSE cross-validation sample. The removed stars are primarily giants with low \teff{} ($<4300$ K) and \logg{} ($<1.5$ dex), which shows low correlation in gradient spectra of \ddpayne{} (see Figure 2 of X19).
We also limit our analysis to stars with correlation coefficient gradient spectra for 
\mgfe{} to larger than 0.65. 
After applying all the selection criteria, we end up with 75 stars physically located in the Galactic bulge. 


Panel (a) and (d) of Figure \ref{f:mgfe_bulge} show the \mgfe{}-\feh{} distribution of stars in MUSE data-cubes towards the Galactic bulge with values measured by \ddpaynea{} model (hereafter MUSE-\ddpaynea{}) with comparison of the same plot using stars from APOGEE-\emph{Payne} \citep{2019ApJ...879...69T}. 
We use the Gaussian kernel estimation and plot the density color map. 
The selection criteria of APOGEE stars is the same as that in \citealt{2019A&A...626A..16R} and the distances were estimated by \citealt{2016A&A...585A..42S}.
\mgfe{} in MUSE-\ddpaynea{} are calibrated by adding an offset. This is because there is a non-negligible \mgfe{}-\teff{} trend in \ddpaynea{} model, which is inherited from the APOGEE-\emph{Payne} values in the training set (See Figure 14 in X19).
The average temperature differences between stars in APOGEE-\emph{Payne} and MUSE-\ddpaynea{} in the bulge will lead to roughly 0.08~dex deviation in \mgfe{}. So we calibrate this deviation here.
This trend is beyond the scope of this study, and more details will be discussed in the Appendix~\ref{app:mg_teff_trend}.
In Figure \ref{f:mgfe_bulge}, both panel (a) and (d) demonstrate two sequences of stars in the bulge: Metal-poor/$\alpha$-rich sequence with center in (\mgfe{}, \feh{}) space at $(0.36,-0.43)$ and metal-rich/$\alpha$-poor sequence at $(0.19, 0.18)$.
These two centers are in agreement in MUSE-\ddpaynea{} and APOGEE-\emph{Payne} samples. In addition, the two sequences merge at \feh{}$\sim-0.07$ dex in both panels. 
All these agreements reassure and validate the precision of the MUSE results, and the ability of \ddpayne{} to measure stellar labels in dense fields.

However, The two plots also show differences.  
The $\alpha$-rich stars in the MUSE-\ddpaynea{} are less dense than those in APOGEE-\emph{Payne} and the distribution is also broader. 
Therefore, the small sample size of MUSE is certainly a limiting factor in comparing the two panels. 
Another important factor that is different in the two panels is the distribution of stars in the $(R_{\mathrm{gc}},z)$ plane (See the top panel of Figure~\ref{f:avrgczgal}), neither is the apparent magnitude selection. MUSE data is restricted to 3 lines of sight, which are all close to the plane.

\subsubsection{\mgfe{}-\feh{} distribution of other regions}
\label{sss:resu-bulge-35}


In the top panel of Figure~\ref{f:avrgczgal}, we can see that in addition to the bulge stars, there are many foreground stars observed in these nine fields. 
We can also study the \mgfe{}-\feh{} distribution of these stars and compare them with distributions in APOGEE-\emph{Payne}. 
Here, we select the stars with \rgc{} between $3\sim5$ and $5\sim7$ kpc, and plot the \mgfe{}-\feh{} distributions in the panel (b), (e), (c), (f) of Figure~\ref{f:mgfe_bulge}, respectively.
This is similar to \citealt{2015ApJ...808..132H, 2021MNRAS.507.5882S}, which used stars in APOGEE DR12 and plotted \alphafe{}-\feh{} distributions in different Galactic locations.
Here the \feh{} and \mgfe{} values in MUSE-\ddpaynea{} are calibrated for the same reason as before.
In this figure, for the same \rgc{}, the distributions of stars in MUSE-\ddpaynea{} and APOGEE-\emph{Payne} samples are in good agreement, with the centers at the same (\mgfe{}, \feh{}) locations.
All the panels are dominated by low-$\alpha$ stars, and the overall trends are consistent.
Despite the small number of MUSE stars (52 in panel (e) and 103 in panel (f)), the good agreement with APOGEE-\emph{Payne} is once again reassuring and validating the precision of the MUSE results.

\section{Estimating exposure times to achieve a given uncertainty}
\label{s:predict}
The uncertainty $\sigma_{\rm X}$ of a stellar label $X$ derived with MUSE using the \ddpayne{} depends on the signal to noise ratio (SNR hereafter) of the spectra, which in turn depends on the exposure time $T_{\rm exp}$ and the magnitude of a star.
The brighter the star and longer exposure time, the higher the SNR and thus the more precise the stellar labels will be.
Therefore, when doing a survey or planning future observations, one should ensure that the exposure time is long enough to reach the expected uncertainty. 
In the next sections, we provide formulas to estimate the $\sigma_{\rm X}$ as a function of $V$ band magnitude and exposure time, to provide a reference for future MUSE observations.
To do this, we first estimate the $\sigma_{\rm X}$ as a function of SNR. Next, we estimate the SNR as a function of magnitude $V$ and $T_{\rm exp}$.

\subsection{Uncertainty of stellar labels as a function of SNR}
\begin{figure}
\includegraphics[width=0.98\columnwidth]{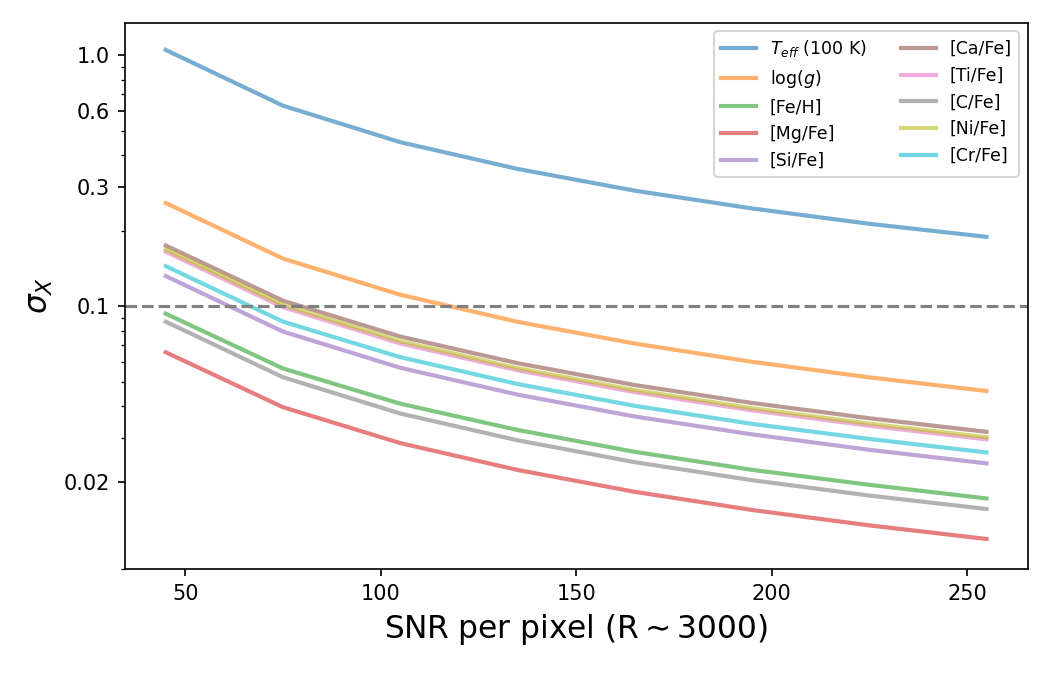}
\caption{
Prediction of precision of stellar labels as a function of SNR. It is derived by estimating the theoretical precision using \crlb{} from typical K-giant spectra from MUSE Exposure Time Calculator in different SNRs. Then a factor is applied to transfer the theoretical precision to real precision for each label which is calculated by Equation~\ref{e:facpair}.}
\label{f:predict_sigma}
\end{figure}

\begin{table}
\caption{Parameters for the equation~\ref{eq:sigmasnr} of uncertainty as a function of SNR.}
    \begin{tabular}{lcclcc}
    \hline
    stellar label & $A_X$ & $B_X$ & stellar label & $A_X$ & $B_X$\\
    \hline
    \teff{} & -0.9875 & 1.652 & \cafe{} & -0.9834 & 0.868\\
    \logg{} & -0.9922 & 1.052 & \tife{} & -0.991 & 0.856\\
    \feh{} & -0.9746 & 0.5823 & \cfe{} & -0.9891 & 0.5737\\
    \mgfe{} & -0.9862 & 0.4486 & \nife{} & -0.9925 & 0.8681\\
    \sife{} & -0.9886 & 0.7546 & \crfe{} & -0.9835 & 0.7855\\
    \hline
    \end{tabular}
\label{tab:tab3}
\end{table}

Figure~\ref{f:predict_sigma} 
shows the uncertainty $\sigma_{X}$ of a stellar label $X$ as a function of SNR. 
We firstly derive theoretical uncertainty $\sigma_{ X,{\rm Theoretical}}$ using \crlb{} \citep{2020ApJS..249...24S} from a typical K-giant spectra. 
SNR is estimated from MUSE Exposure Time Calculator\footnote{\url{https://www.eso.org/observing/etc/bin/gen/form?INS.NAME=MUSE+INS.MODE=swspectr}}.
The air-mass and seeing are set as 1.5 and $0.8^{\prime \prime}$. 
Then $\sigma_{X}$ is calculated by
\begin{equation}
\sigma_{X}=\sigma_{ X,{\rm Theoretical}}({\rm SNR})f_{X,{\rm corr}},
\end{equation}
where $f_{X,\rm corr}$ is the scale factor applied to transform the theoretical uncertainty to the observed uncertainty for each label $X$. 
The factor is calculated from MUSE repeat observations by computing the dispersion of pairwise differences as follows.
\begin{equation}
f_{X,\rm corr} = \text{std-dev} \left( \frac{X_{m}-X_{n}}{
\sigma^{2}_{X,{\rm Theoretical},m}
+\sigma^{2}_{X,{\rm Theoretical},n}} \right),
\label{e:facpair}
\end{equation}
where $m$ and $n$ represent the repeated observations of a  star. $\sigma_{X,{\rm Theoretical}}$ is the theoretical uncertainty of a spectra as given by \crlb{}. 
Comparing with the error from \ddpayne{}, \crlb{} estimates are normally 1.5 times smaller (see Fig. 21 in \citealt{2020ApJS..249...24S}), so we multiply this factor.
Then $\sigma_{ X,{\rm Theoretical}}$ mentioned in this section means the theoretical uncertainty of multiplying 1.5.
Figure~\ref{f:predict_sigma} indicates that the uncertainty will decrease as the SNR increases.
Generally, \mgfe{} has the lowest uncertainty with $\sigma_{\feh}<0.1$ dex for SNR higher than 29.44 \perpixel{}.
As for other elements, to make the uncertainty less than 0.1 dex, the SNR needs to be higher than $\sim$39.00, 42.02, 59.54, 79.35, 74.62, 76.24, 65.39 \perpixel{} for \cfe{}, \feh{}, \sife{}, \cafe{}, \tife{}, \nife{} and \crfe{}, respectively.
To facilitate the calculation of the SNR required to achieve a given uncertainty, we provide analytical fits to the uncertainty curves shown in Figure~\ref{f:predict_sigma}.
The curves are fitted with the functional form of 
\begin{equation}
\log \sigma_{X} = A_X \log ({\rm SNR}) + B_X,
\label{eq:sigmasnr}
\end{equation}
The coefficients $A_X$ and $B_X$ for each label are listed in Table~\ref{tab:tab3}.

\subsection{SNR as a function of magnitude and exposure time}

\begin{figure}
\includegraphics[width=0.98\columnwidth]{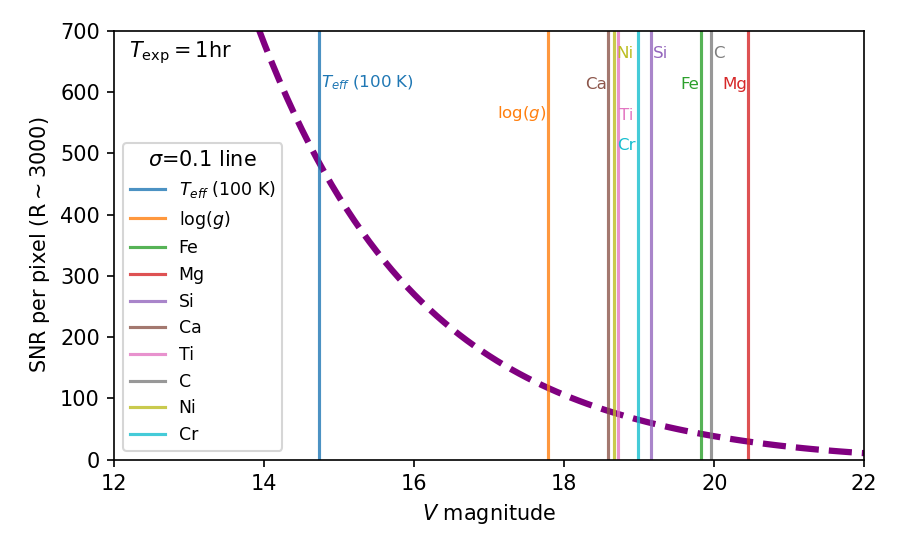}
\caption{
SNR of MUSE spectra at $R\sim3000$ as a function of $V$ magnitude (purple line) for an exposure time of one hour. Each colored line represents the threshold where the uncertainty of a label reaches 0.1 dex (or 10 K),
which is derived using equation~\ref{eq:sigmasnr}. 
All chemical abundances can be derived with uncertainty less than 0.1 dex till $V=18.5$ mag, for \mgfe{}, \cfe{} and \feh{} even till $V=19.7$ mag.
}
\label{f:mag_snr}
\end{figure}

Using equation~\ref{eq:sigmasnr}, with the known SNR we can obtain the uncertainty of stellar labels from its spectra. 
We know SNR depends on magnitude $V$, exposure time $T_{\exp}$, and air-mass.
Therefore, when making science goals, what we care about is how long the exposure time is needed to achieve the required label uncertainty for stars with different magnitudes.
Here we provide an analytical function to calculate SNR with given $V$ magnitude and exposure time $T_{\exp}$ as 
\begin{equation}
{\rm SNR}={\rm SNR}_{\rm 1 hr}(V) \sqrt{\left(T_{\exp } / 3600 \mathrm{s}\right)}.
\label{eq:snrsnr1rexp}
\end{equation}
Here ${\rm SNR}_{\rm 1 hr}(V)$ is the SNR estimated from the K-giant spectra by the MUSE Exposure Time Calculator for an exposure of one hour.
The analytical function is given by
\begin{equation}
{\rm SNR}_{\rm 1 hr}(V)=\frac{462.34 \times 10^{0.4(14.83-V)}}{\sqrt{0.00223+10^{0.4(14.83-V)}}},
\label{eq:snr1hv}
\end{equation}
This idea is taken from \citealt{2018MNRAS.473.2004S}. For the above equations, we set the air-mass as 1.5 and seeing as $0.8^{\prime \prime}$. Therefore, by combining equations~\ref{eq:sigmasnr}-\ref{eq:snr1hv}, with a given magnitude $V$ and exposure time $T_{\exp}$, one can predict the uncertainty of each label.

Figure~\ref{f:mag_snr} provides the SNR as a function of $V$ magnitude for an one-hour exposure (purple line). 
We also plot the threshold of maximum magnitude for each label where the uncertainty reaches 0.1 dex or 10K, which are derived by  Figure~\ref{f:predict_sigma}.
As we can see, all chemical abundances can be derived with uncertainty less than 0.1 dex till $V=18.5$ mag, for \mgfe{}, \cfe{} and \feh{} even till $V=19.7$ mag.
This figure combining with equations~\ref{eq:sigmasnr}-\ref{eq:snr1hv} can be used as a guide for future research, providing an indication of the exposure time for different magnitude stars to achieve the expected uncertainty.

\section{Discussions}
\label{s:discuss}
\subsection{Performance of MUSE stellar label extraction under different scenarios}

Ideally, the methods outlined in this paper to determine stellar labels would be applicable to all stars in MUSE data-cubes. 
However, the label ranges of the \ddpayne{} training set do not cover the space for all potential observations: for stars outside of this parameter space, for example having very low metallicity $(\feh<-1.5)$ or very high effective temperatures $(\teff>7000K)$, labels are determined through extrapolation of the model.
Despite this, some studies have shown \ddpayne{} can do some moderate extrapolation in the linear region of the gradient spectra varying with labels. 
This is because \ddpayne{} imposes the gradient spectra to be similar to that of the Kurucz models \citep{1970SAOSR.309.....K, 1993KurCD..18.....K, 2005MSAIS...8...14K}, which is an advantage of such a hybrid method compared to pure data-driven models.
For example, the LAMOST \feh{} estimates can be robust down to $-2.5$ dex (see Fig. 10 of \citealt{2019ApJ...887..237C}); \ddpayne{} can also generate reasonable [Ba/Fe] stars with $1<{\rm [Ba/Fe]}<3$ dex and $\teff{}>7000$K \citep{2020ApJ...898...28X}.

Typically in MUSE data, stellar spectra come from the following types of fields: random individual stars near a galaxy, high extinction fields, and globular clusters.
In each of these scenarios the 
performance of the spectroscopic analysis is different and some of them pose unique challenges for analysis.

For individual stars near a galaxy, they typically have high SNR, as observations in galaxy fields generally have long exposure times. Therefore, as long as the PSF does not overlap with the nearby galaxy, it is relatively simple to extract and estimate the stellar labels. All of the common stars between MUSE and LAMOST belong to this case; one typical spectrum is shown in Figure~\ref{f:muse_ddp_comparison}.
For these stars, stellar labels will have high precision due to their high SNR, as well as the fact that most of them belong to the nearby disk, which are stellar populations that are well covered by the \ddpayne{} training sets.

For stars in high extinction fields, such as the bulge, the spectra are restricted to a narrower range in wavelength and the SNR of the extracted spectra is in general low (due to a combination of large distance and dust).  For low SNR, in addition to SNR directly affecting precision, the emission lines are comparatively stronger and more difficult to remove. The above-listed  factors lead to lower precision in stellar labels.
Hence, in high extinction regions, one can only study giants which are intrinsically bright. 
Therefore, we can only get reliable label estimates for stars with relatively low extinction ($A_V<6$ mag) along the bulge lines of sight, which was shown in the middle plot of Figure~\ref{f:avrgczgal}.
As for stars of $A_V>6$ mag in the Galactic center, the dust will make the giants difficult to observe with typical MUSE exposure times.

For globular clusters, which have high stellar density, our work will be of particular importance.
It will provide an automated way to determine high precision stellar labels for these dense fields down to the center of the cluster; previous studies using fiber-fed spectrographs were often limited to the outskirts of these objects.
Additionally, it is possible to increase the precision of stellar parameters and several abundance estimations by making use of full wavelength fitting, which has not been done so far.
Previously, the MUSE GC survey \citep{2016A&A...588A.148H, 2016A&A...588A.149K} observed 26 globular star clusters, with most of these studies focusing on specific features of spectra or just metallicity, e.g. emission-line studies \citep{2019A&A...631A.118G}, the dynamics using CaT absorption lines \citep{2018MNRAS.473.5591K}, the relative abundance variations using the equivalent widths \citep{2019A&A...631A..14L}, metallicity distributions \citep{2020A&A...635A.114H, 2021A&A...653L...8L}. 
Our method has the potential to enable more chemical abundances measured automatically by the full-wavelength fitting frameworks, rather than applying any spectral library to execute the fittings.
Even though globular clusters typically have $\feh{}<-1.0$~dex, some of them are still in the parameter range of \ddpayne{}, and the moderate linear extrapolation ability of \ddpayne{} still enables this method to measure robust stellar labels. We will perform the test of it in the future work.

\subsection{Computational challenges in spectral extraction}

In this work, the spectral extraction processes are executed using PampelMUSE \citep{2013A&A...549A..71K}. The typical time spent on extracting spectra from each data-cube with a 24-core CPU and 32GB RAM is $\sim30$ minutes and it is independent of the number of stars in the field. Therefore, for the verification data analyzed in this paper (the cross-match of LAMOST and MUSE), we spent more time on spectral extraction for these cubes compared to the data-cubes in the bulge, even though there were two orders of magnitude more spectra extracted from the bulge fields. Hence, to perform a survey of all individual stars in publicly available MUSE data-cubes, the most time-consuming step will be the extraction of spectra.

For individual stars in sparse fields or more limited studies focused purely on kinematics in dense fields, it is also possible to use other stellar extraction strategies such as aperture photometry in IRAF for each layer subtracted by the sky estimated locally (e.g. \citealt{2018A&A...616A..83V}). 
This approach has the advantage of retaining as many stars as possible and has a more accurate estimate of the sky background, which will lead to reduced artifacts in the final spectra \citep{2018A&A...616A..83V}.
For regions with modest crowding (i.e., $<200$ stars per field), aperture photometry and PSF-fitting could yield similar results. However, PSF-fitting is still necessary when dealing with more crowded fields ($>200$ stars per field) as there will be significant blending for stars in these dense regions. 
\citealt{2013A&A...549A..71K} tested the performance of \pampelmuse{} PSF-fitting using a simulated MUSE data-cube towards the center of a globular cluster 47 Tuc. Under the average seeing of 0.8 arcsec, $\sim$5000 useful spectra could be extracted. In good conditions, this value can be 3 or 4 times larger. With severe crowding, aperture photometry will suffer from significant blending between stellar spectra. Therefore, for globular clusters, we still recommend PampelMUSE.

\subsection{Future research}

There are several aspects in the data pipeline that can be improved in the future.
In the methods outlined in this paper, stellar labels are estimated from the \ddpayne{}, which depends sensitively on the precision of labels in GALAH DR2 \citep{2018MNRAS.478.4513B} and APOGEE-\emph{Payne} \citep{2019ApJ...879...69T}. 
Given that improved versions of both these catalogs are now available, GALAH DR3 \citep{2021MNRAS.506..150B} and APOGEE DR16 \citep{2020ApJS..249....3A}, one needs to update the \ddpayne{} models.
In addition, we can train a new \ddpayne{} model if a sufficient number of common stars in MUSE and other high-resolution spectroscopic surveys for a training set is available.
The advantage is that there will be no need to degrade the MUSE spectra to a lower resolution, which will help extraction of blended features.
In addition, some absorption lines such as [Na/Fe] and [O/Fe] are masked in the LAMOST spectra due to dichroic issues in the wavelength range $(5803-5966\Angstrom)$.
This range can now be used when working in MUSE non-AO mode.
It will enable to estimate precise sodium and oxygen abundances, which is essential in identifying different stellar populations in globular clusters \citep{2018ARA&A..56...83B}. The primary difficulty for retraining is having a training set that is both large and covers a wide range of parameter space.

In the future, the method outlined here can be utilized for many purposes. A survey can be executed based on stars from publicly available MUSE data-cubes using Gaia eDR3 \citep{2021A&A...649A...1G} and SkyMapper \citep{2007PASA...24....1K} as photometric input catalogs. Our method can be applied to estimate their stellar labels and compile a catalog that will be useful for Galactic archaeology studies.
For the Galactic bulge, our method has the opportunity to study the chemistry and dynamics of the Galactic center in greater detail. 
However, more observational data is needed, such as fields in Baade's window, which are less affected by the dust.
In addition, this method can be applied to globular clusters observed by \citealt{2016A&A...588A.148H, 2016A&A...588A.149K} to study their chemistry, and identify multiple stellar populations if any, and their connection to kinematics.
Moreover, the interest in using MUSE for spectroscopy of resolved stellar populations has been growing quite strongly, even beyond the limits of the Galaxy (e.g. \citealt{2019AN....340..989R}), our method can also provide parameter estimations for these targets.

With the launch of the BlueMUSE project \citep{2019arXiv190601657R}, we will obtain spectra in the wavelength range between $3500-5800\Angstrom$, which is also covered by LAMOST. As mentioned previously, the loss of the blue portion of the spectrum in MUSE is the primary reason why we are not able to estimate as many precise chemical elements as those determined from LAMOST spectra.
BlueMUSE will provide the opportunity to measure these abundances with higher precision, and for more elements that are currently not available; critically the s-process elements such as [Y/Fe] and [Ba/Fe] which provide key age diagnostics for stellar targets \citep{2020arXiv201113745H}. 
The blue portion will also bring strong absorption features for determining precise abundances of Na, C, N, and O, which are essential for identifying multiple populations in globular clusters. 
In addition, MAVIS \citep{2020arXiv200909242M}, which is the new IFS instrument being built for ESO's VLT AOF (Adaptive Optics Facility, UT4 Yepun) also covers a wavelength range $3700-10000\Angstrom$ similar to BlueMUSE and MUSE.  With the lower spatial sampling of $0.02^{\prime \prime} - 0.05^{\prime \prime}$, we have the opportunity to study the core of globular clusters with more stars providing an insight into its chemistry and dynamics. Because of the similar wavelength coverage of \ddpayne{} and these instruments, our method can be directly applied to them.

Finally, based on the fact that the study of resolved stellar populations represents one of the major science cases for the European Extremely Large Telescope (ELT) \citep{2007Msngr.127...11G, 2009ASSP....9..225H}, it is reassuring that our method is paving the way for future studies in multi-dimensional chemical space.
For the Galactic bulge, ELT will allow us to measure chemical abundances all the way down to the main sequence turn-off stars  \citep{Minniti_2007}. For these targets, we can use BSTEP \citep{2018MNRAS.473.2004S} to map the age distribution and reveal the detailed formation and evolution history of the innermost bulge, the peanut bar, and the long bar. 

For Local Group galaxies, the ELT can also derive more stellar spectra of bright main-sequence (potentially even turn-off) stars and giants with necessary SNR than VLT (e.g. \citealt{2019MNRAS.486.5263M, 2020AJ....159..152C}); this includes the SMC and LMC, as well as more distant galaxies such as Andromeda and M33, extending Galactic Archaeology beyond the Milky Way to the Local Group. 
With the unprecedented detection sensitivity of ELT, our method can provide a way for future stellar label estimation and science goals for the ELT-related IFS instruments, to measure stellar labels of stars beyond the Local Group \citep{Wyse_2005, 2008RMxAC..33...23E, 2012PASP..124..653G}.

\section{Summary}
\label{s:summary}
In this study, we applied \ddpayne{} \citep{2019ApJS..245...34X}, which was developed for LAMOST spectra, on MUSE data-cubes and estimated \teff{}, \logg{}, \feh{} and chemical abundances of Mg, Si, Ti, Ca for individual stars in MUSE data-cubes. 

By comparing our MUSE results with that of LAMOST using common stars, we show that despite instrumental differences, it is possible to extract precise stellar parameters and abundances with typical dispersion of \teff{}, \logg{} and other chemical abundances less than 75K and 0.15 dex and 0.1 dex, respectively.

In dense fields, which is the unique advantage of the MUSE instrument, we selected 29 data-cubes towards the Galactic center.
Based on previous studies \citep{2019A&A...626A..16R, 2015ApJ...808..132H, 2021MNRAS.507.5882S}, by comparing the \feh{}-\mgfe{} distribution estimated by our method in the bulge and inner disk region with stars from APOGEE-\emph{Payne}, we found excellent agreements in two sequences with similar center values and overall trends.
This indicates the ability of our method to measure chemical abundances in regions with high stellar density and extinction. 
In addition, we also studies the extinction as a function of \rgc{} in the direction of the Galactic center and compared these results with the prediction from \textit{Galaxia} \citep{2011ApJ...730....3S}. 
We found there is a qualitative agreement with the overall trends, but the scale height of the dust predicted by \textit{Galaxia} is  higher than observations.

In addition, we provided analytical formulas to predict the precision of each label as a function of $V$ magnitude and exposure time. This can be used for observational proposal designing before a survey to achieve the expected label uncertainty.

In the future, this method can be applied to do a survey for all individual stars in public MUSE cubes and estimate their stellar labels. The result can be a supplementary catalog for all-sky spectroscopic surveys. 
For the Galactic bulge, more observations are needed to expand the sample size. By combining the chemical and kinematic information, there is a great opportunity to study the evolution and assembly history of the Galactic center. 
This method can also be applied to MUSE observations on globular clusters, to study the connection between kinematics and multiple populations.

This is the first time to measure stellar labels using full wavelength fitting in dense fields observed by the MUSE instrument. We demonstrate that despite PSF/LSF being different, it is possible to apply a model trained on spectra from one instrument to another. Given a large number of spectrographs having wavelength similar to LAMOST, our results suggest that one can easily do detailed spectroscopic analysis with them. In the future, this method can be used directly for BlueMUSE \citep{2019arXiv190601657R} and MAVIS \citep{2020arXiv200909242M}. It will also help with the label measurement development for IFS instruments on ELT \citep{2007Msngr.127...11G}, to study Galactic archaeology from the Milky Way to somewhere more than 10 Mpc away.

\section*{Acknowledgements}
We thank Xu Zhang, David M. Nataf, Thorsten Tepper Garcia, Sven Buder, and Jesse Van de Sande for their very useful comments.

This research has made use of the services of the ESO Science Archive Facility. The results are based on public data released from the MUSE commissioning observations at the VLT Yepun (UT4) telescope.
This work has also made use of the data from LAMOST. The Large Sky Area Multi-Object Fiber Spectroscopic Telescope (LAMOST) is a National Major Scientific Project built by the Chinese Academy of Sciences. Funding for the project has been provided by the National Development and Reform Commission. LAMOST is operated and managed by the National Astronomical Observatories, Chinese Academy of Sciences.
This work was also based on data products from VVV Survey observations made with the VISTA telescope at the ESO Paranal Observatory under programme ID 179.B-2002.

This research has made use of the VizieR catalogue access tool, CDS, Strasbourg, France (DOI: 10.26093/cds/vizier). The original description of the VizieR service was published in \citealt{2000A&AS..143...33B}; Astropy\footnote{\url{http://www.astropy.org}}, a community-developed core Python package for Astronomy \citep{2013A&A...558A..33A, 2018AJ....156..123A}, numpy \citep{2020Natur.585..357H}, scipy \citep{2020NatMe..17..261V}, matplotlib \citep{2007CSE.....9...90H}, ZAP \citep{2016MNRAS.458.3210S}, PampelMUSE \citep{2013A&A...549A..71K}, DD-Payne \citep{2019ApJS..245...34X} and MPDAF \citep{2017arXiv171003554P}.

We acknowledge the University of Sydney HPC service at The University of Sydney for providing HPC resources that have contributed to the research results reported in this paper.

ZW is supported by the China Scholarship Council and Australian Research Council Centre of Excellence for All Sky Astrophysics in Three Dimensions (ASTRO-3D) through project number CE170100013. 
YST is grateful to be supported by the NASA Hubble Fellowship grant HST-HF2-51425.001 awarded by the Space Telescope Science Institute. YST acknowledges financial support from the Australian Research Council through DECRA Fellowship DE220101520.

\section*{Data Availability}

The data underlying this article will be shared on reasonable request to the corresponding author.



\bibliographystyle{mnras}
\bibliography{biblio.bib} 




\appendix

\section{The [Mg/Fe]-Temperature trend in the model}
\label{app:mg_teff_trend}

\begin{figure*}
\includegraphics[width=2.0\columnwidth]{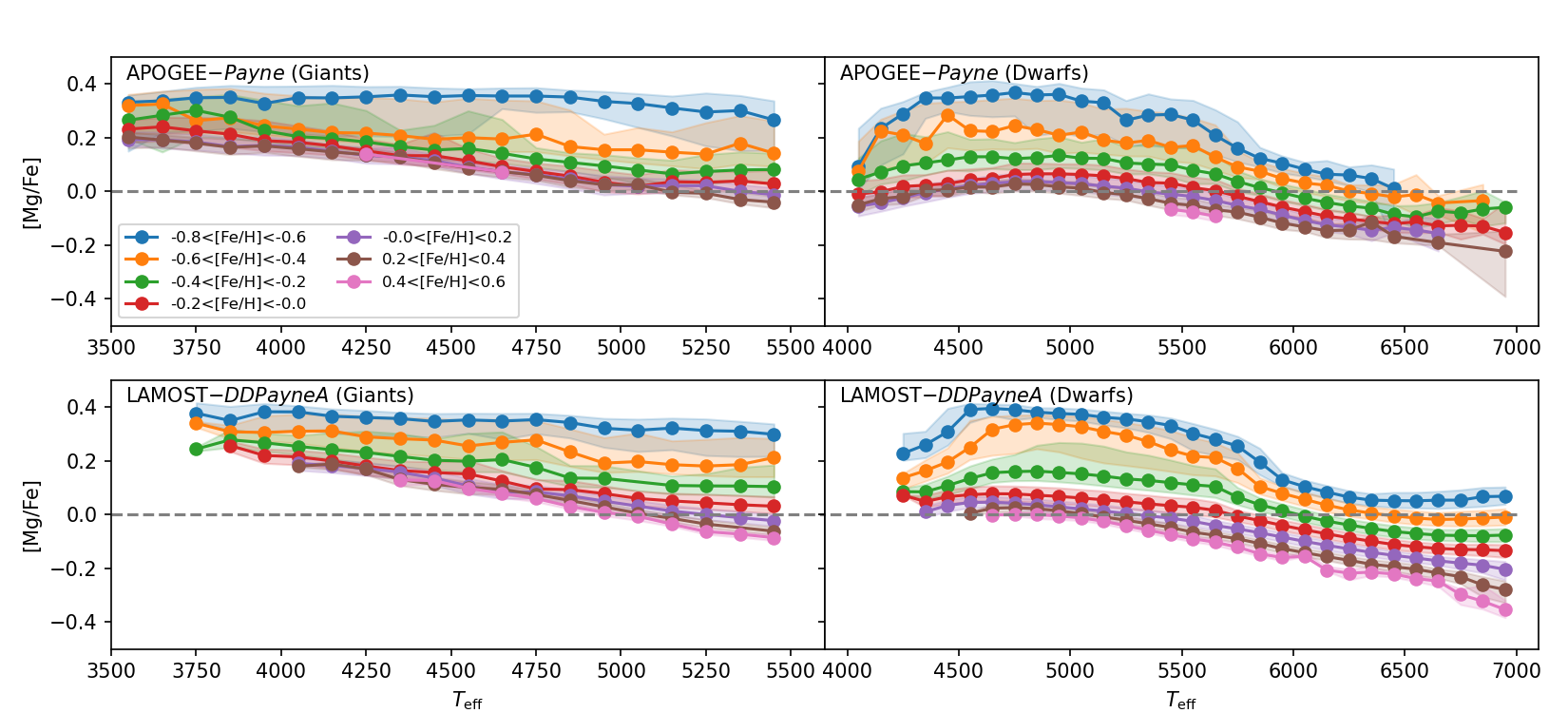}
\caption{
Stellar \mgfe{} as a function of \teff{} for stars in different metallicity bins. 
The top row shows the results from APOGEE-\emph{Payne} \citep{2019ApJ...879...69T}, the the bottom row the LAMOST-\ddpaynea{} results (X19). The dots and filled ranges illustrate the median values and standard deviations at different \teff{} bins. Both of the results demonstrate that there is a \teff{}-\mgfe{} trend in all the metallicity bins. The slopes of these trends are illustrated in Table~\ref{tab:app_tab1}. In addition, the trends in APOGEE-\emph{Payne} and LAMOST-\ddpaynea{} are consistent because \ddpaynea{} is trained from labels in APOGEE-\emph{Payne} and inherited its features. 
}
\label{f:appen-mg-teff-trend}
\end{figure*}


\begin{figure*}
\includegraphics[width=2\columnwidth]{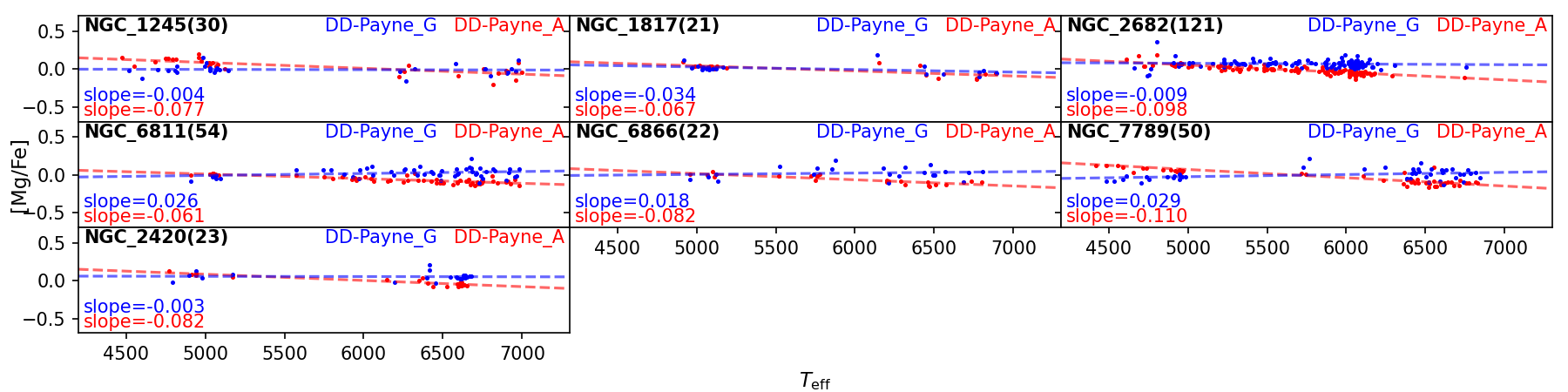}
\caption{
Stellar \mgfe{} as a function of \teff{} for stars in seven open clusters (the name and stellar number are marked in each panel). 
The stars are from LAMOST DR5 and selected by the criteria in \citealt{2020A&A...640A.127Z}. Their \mgfe{} and \teff{} are then measured by the \ddpayne{} models. 
The results measured by \ddpaynea{} show clear \mgfe{}-\teff{} trends with larger slopes than \ddpayneg{}. Since the metallicity and abundances of stars in open clusters are homogeneous, which means no galactic evolution or selection function applied, the trends illustrated here are caused by labels in APOGEE-\emph{Payne}.
}
\label{f:appen_trend_oc}
\end{figure*}

\begin{figure*}
\includegraphics[width=1.95\columnwidth]{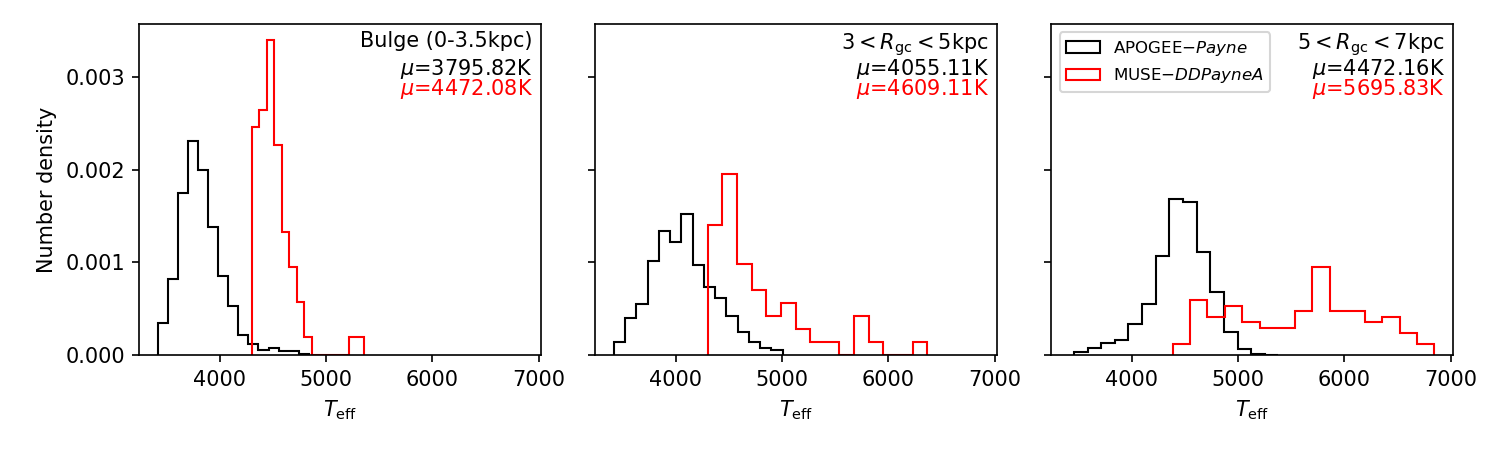}
\caption{
\teff{} density histograms of stars in Figure~\ref{f:mgfe_bulge} as a function of \rgc{}.
Results from APOGEE-\emph{Payne} \citep{2019ApJ...879...69T} are illustrated in black, and the red shows the MUSE-\ddpaynea{} results. The medians ($\mu$) are marked on each panel. This figure shows the difference of \teff{} in the APOGEE-\emph{Payne} and MUSE-\ddpaynea{} samples, which can lead to a shift in \mgfe{}. 
}
\label{f:appen_teff_diff}
\end{figure*}

The \mgfe{}-\teff{} trends was found in the \ddpayne{} model in X19 (see Figure~14), where for dwarfs with $-0.2<\feh<0.2$, \mgfe{} measured by \ddpaynea{} decreases with \teff{}. 
In addition to \mgfe{}, some other abundances also show similar non-negligible trends. X19 verified that these trends are inherited from the APOGEE-\emph{Payne} values in the training set. 
Since no abundance calibration was applied to APOGEE-\emph{Payne} values \citep{2019ApJ...879...69T}, the \mgfe{}-\teff{} trend should come from the neural network model \thepayne{}, i.e., stellar astrophysics.

To investigate this trend, we take APOGEE-\emph{Payne} \citep{2019ApJ...879...69T} and LAMOST-\ddpaynea{} (X19) catalogs, separate stars into giants and dwarfs and divide them into different \feh{} bins, then plot the \mgfe{}-\teff{} distributions in Figure~\ref{f:appen-mg-teff-trend}. 
Selection criteria are applied to these two samples to discard stars with bad fittings: For APOGEE-\emph{Payne}, excluding stars with any of the flags: "quality"$=$"bad" and $\chi^2>20$; 
For LAMOST-\ddpaynea{}, excluding stars with any of the flags: $\rm SNR_{g}<30$, $\chi^2>20$, "qflag-chi2"$=$"bad" and "flag-singlestar"$=$"no".
The dots and filled ranges illustrate the median values and the standard deviations at different \teff{} bins. 
In this Figure, both the giants and dwarfs demonstrate that there is a clear \teff{}-\mgfe{} trend in all the metallicity bins. The slopes of these trends (illustrated in Table~\ref{tab:app_tab1}) in the two catalogs are similar as well.


Another reason leading to this trend is the Galactic evolution and selection functions, i.e., stars with higher \teff{} tend to be younger, and therefore have lower \mgfe{} than stars with lower \teff{}. Even though the metallicity bins in Figure~\ref{f:appen-mg-teff-trend} are chosen to be small, it is not sufficient to conclude the \mgfe{}-\teff{} trend is all caused by the model. 
To explore the proportion of this trend due to the model, we select seven open clusters from LAMOST DR5 and plot \mgfe{} as a function of \teff{} in Figure~\ref{f:appen_trend_oc}, where the name and star number are marked in each panel. 
The stars in the open clusters are selected by the criteria in \citealt{2020A&A...640A.127Z}. Their \mgfe{} and \teff{} are then measured by the \ddpayne{} models. 
In all the panels, results from \ddpaynea{} show clear \mgfe{}-\teff{} trends with larger slopes than those from \ddpayneg{}, which is $-0.082$~dex/1000K on average.
Since open clusters are found to be chemical homogeneous \citep{2016ApJ...817...49B, 2018ApJ...853..198N}, which means no galactic evolution or selection function is applied, the trend illustrated here is verified to be caused by the model.
Combining with the slopes in Figure~\ref{f:appen-mg-teff-trend}, for stars with $\feh{}\sim0$, which is the approximate \feh{} for these open clusters, the average slope is $-0.124$~dex/1000K. 
Therefore, the contribution of this trend is comparable but higher from the model (stellar astrophysics).


In APOGEE-\emph{Payne} \citep{2019ApJ...879...69T}, the authors chose to not calibrate the abundance trend because at the \teff{} of APOGEE range (4000K-5000K), it was shown that an ab-initio fitting of the spectra will not incur much [Mg/H]-\teff{} trend (see their Figure 15).
However, we found that even with the ab-initio fitting, for stars with a wider \teff{} range, in the \mgfe{} scale, the values in APOGEE-\emph{Payne} do have a non-negligible trend, which is similar to the APOGEE pipeline ASPCAP \citep{2015AJ....150..148H} and requires calibration.

Therefore, when plotting the \mgfe{}-\feh{} distributions, 
additional calibration should be added to eliminate this impact. 
In Figure~\ref{f:appen_teff_diff}, we plot the \teff{} density histograms of stars in Figure~\ref{f:mgfe_bulge} in different \rgc{} regions.
Results from APOGEE-\emph{Payne} are illustrated in black, and the red shows the MUSE-\ddpaynea{} results. The medians ($\mu$) are marked on each panel. This figure illustrates the difference of \teff{} in the APOGEE-\emph{Payne} and MUSE-\ddpaynea{} samples. 
Stars with $5<R_{\rm gc}<7$ kpc in MUSE-\ddpaynea{} are mostly dwarfs, and the other samples are mostly giants.
Referring to Figure~\ref{f:appen-mg-teff-trend}, we can roughly estimate that for stars in the bulge, $3<R_{\rm gc}<5$ and $5<R_{\rm gc}<7$ kpc, the deviations of \mgfe{} between two samples are $\sim0.08, 0.06$ and $0.08$~dex, respectively. These are the values added to \mgfe{} in Figure~\ref{f:mgfe_bulge} for calibrating.


\begin{table}
\caption{Slope for the \mgfe{}-\teff{} trend (dex/1000K) in different metallicity bins for stars in APOGEE-\emph{Payne} and LAMOST-\ddpaynea{}.}
    \begin{tabular}{*5c}
    \hline
    \multirow{2}{*}{\feh{} bins} & \multicolumn{2}{c}{APOGEE-\emph{Payne}} & \multicolumn{2}{c}{LAMOST-\ddpaynea{}} \\
    & giants & dwarfs & giants & dwarfs \\
    \hline
    $(-0.8, -0.6)$ & -0.025 & -0.117 & -0.044 & -0.141 \\
    $(-0.6, -0.4)$ & -0.086 & -0.105 & -0.092 & -0.132 \\
    $(-0.4, -0.2)$ & -0.125 & -0.080 & -0.111 & -0.098 \\
    $(-0.2, 0.0)$ & -0.128 & -0.071 & -0.142 & -0.093 \\
    $(0.0, 0.2)$ & -0.118 & -0.064 & -0.162 & -0.104 \\
    $(0.2, 0.4)$ & -0.131 & -0.078 & -0.186 & -0.132 \\
    $(0.4, 0.6)$ & -0.165 & -0.130 & -0.212 & -0.154 \\
    \hline
    \end{tabular}
\label{tab:app_tab1}
\end{table}


\bsp	
\label{lastpage}
\end{document}